 \documentclass[final,3p,times,twocolumn]{elsarticle}

\biboptions{numbers,sort&compress}
\usepackage{url}
\usepackage{longtable,rotating,multirow}
\usepackage{setspace}
\usepackage{verbatim}

\usepackage{graphicx}      
\usepackage{natbib}        
\usepackage{amsmath}
\usepackage{amsfonts,amssymb}
\usepackage{bm}
\usepackage{threeparttable}

\usepackage{lineno}

\journal{Automatica}

\newtheorem{theorem}{Theorem}
\newtheorem{corollary}{Corollary}
\newtheorem{definition}{Definition}
\newtheorem{lemma}{Lemma}
\newtheorem{proposition}{Proposition}
\newtheorem{assumption}{Assumption}
\newtheorem{remark}{Remark}
\newtheorem{problem}{Problem}
\begin{document}

\begin{frontmatter}



\title{Detection of Intermittent Faults Based on an Optimally Weighted Moving Average $T^2$ Control Chart with Stationary Observations}


\author[Beijing,UMD]{Yinghong~Zhao}\ead{zyh14@mails.tsinghua.edu.cn}
\author[Beijing]{Xiao~He}\ead{hexiao@mail.tsinghua.edu.cn}
\author[Beijing]{Junfeng~Zhang}\ead{jf-zhang13@mails.tsinghua.edu.cn}
\author[Qingdao]{Hongquan~Ji}\ead{jihongquansd@126.com}
\author[Qingdao,Beijing]{Donghua~Zhou\corref{cor}}\ead{zdh@mail.tsinghua.edu.cn}
\author[UMD]{Michael~G.~Pecht}\ead{pecht@umd.edu}
\address[Beijing]{Department of Automation, Beijing National Research Center for Information Science and Technology (BNRist), Tsinghua University, Beijing 100084, China}
\address[Qingdao]{College of Electrical Engineering and Automation, Shandong University of Science and Technology, Qingdao 266590, China}
\address[UMD]{Center for Advanced Life Cycle Engineering (CALCE), University of Maryland, College Park, MD 20742, USA}
\cortext[cor]{This work was supported by the National Natural Science Foundation of China (NSFC) under Grants 61751307, 61733009, the Research Fund for the Taishan Scholar Project of Shandong Province of China (LZB2015-162), and the Key Project from Natural Sciences Foundation of Guangdong Province under Grant 2018B030311054. Corresponding author: Donghua~Zhou.}

\begin{abstract}
The moving average (MA)-type scheme, also known as the smoothing method, has been well established within the multivariate statistical process monitoring (MSPM) framework since the 1990s. However, its theoretical basis is still limited to smoothing independent data, and the optimality of its equally or exponentially weighted scheme remains unproven. This paper aims to weaken the independence assumption in the existing MA method, and then extend it to a broader area of dealing with autocorrelated weakly stationary processes.
With the discovery of the non-optimality of the equally and exponentially weighted schemes used for fault detection when data have autocorrelation, the essence that they do not effectively utilize the correlation information of samples is revealed, giving birth to an optimally weighted moving average (OWMA) theory.
The OWMA method is combined with the Hotelling's $T^2$ statistic to form an OWMA $T^2$ control chart (OWMA-TCC), in order to detect a more challenging type of fault, i.e., intermittent fault (IF). Different from the MA scheme that puts an equal weight on samples within a time window, OWMA-TCC uses correlation (autocorrelation and cross-correlation) information to find an optimal weight vector (OWV) for the purpose of IF detection (IFD).
In order to achieve a best IFD performance, the concept of IF detectability is defined and corresponding detectability conditions are provided, which further serve as selection criteria of the OWV. Then, the OWV is given in the form of a solution to nonlinear equations, whose existence is proven with the aid of the Brouwer fixed-point theory.
Moreover, symmetrical structure of the OWV is revealed, and the optimality of the MA scheme for any IF directions when data exhibit no autocorrelation is proven.
Finally, simulations on a numerical example and the continuous stirred tank reactor process are carried out to give a comprehensive comparison among OWMA-TCC and several existing static and dynamic MSPM methods. The results show a superior IFD performance of the developed methods.
\end{abstract}


\end{frontmatter}


\section{Introduction}\label{IntroductionSec}
Fault detection (FD) for industrial processes with multivariate statistical process monitoring (MSPM) methods has been a hot topic in the past few decades \cite{Choudhury2004Diagnosis,Qin2003Statistical,Kruger2004Improved}.
MSPM methods use various control charts to check statistical properties of process variables, among which $T^2$ control chart is one of the most effective ones since the Hotelling's $T^2$ statistic is admissible and most powerful in certain classes of hypothesis tests \cite{Wierda1994Multivariate}. Permanent faults (PFs) are serious threats to the system reliability, since once PFs occur, they take effect permanently unless removed by external intervention. In practice, many kinds of PFs evolve gradually from intermittent faults (IFs) \cite{Zhou2020Review}. This implies that if faults are detected in this early stage, severe damage caused by PFs, such as system disruptions, plant shutdowns and even safety accidents, can be effectively avoided. In this regard, the detection of IFs is an important means to improve the system reliability. IFs have been recently of noticeable interest, and thus a review of their current research status has been published \cite{Zhou2020Review}.

The IF is a kind of non-permanent fault that lasts a limited period of time and then disappears without any treatment \cite{Zhou2020Review}. So far, the IF detection (IFD) problem has been investigated under both model-based \cite{Carvalho2017Diagnosability,Zhang2020Robust} and data-driven frameworks. As for data-driven methods, several powerful tools such as the signal analysis \cite{Obeid2017Early}, decision forest \cite{Singh2012Decision}, dynamic Bayesian network \cite{Cai2017A} and MSPM \cite{Monekosso2013Data} methods have been utilized to detect IFs in different application domains. Signal analysis methods are suitable to process unidimensional signals that possess periodicity. Decision forest and dynamic Bayesian network methods can take fully use of the historical data of various faults. MSPM methods are easy to handle high-dimensional and correlated variables, and historical fault data are not necessary.

Among various MSPM methods \cite{Monekosso2013Data,Du2017Comparison,Du2018Fault,Niu2019Fault,Zhao2018Intermittent,Zhao2019Detecting,Li2016Fault,Bakdi2017Anew,Bakdi2017Fault,Bakdi2019Nonparametric,Kammammettu2019Change}, principal component analysis (PCA) and canonical correlation analysis (CCA) were used for detecting intermittent sensor faults caused by electrical interference in a smart home sensor network \cite{Monekosso2013Data}. The Gaussian Process (GP) method, which is a Bayesian non-parametric supervised learning approach, has been recently in widespread use for many regression and classification tasks. In \cite{Du2017Comparison}, a GP regression (GPR) model was established to estimate the mean and variance of the occurring intermittent stochastic faults with available measurements, and to detect the IFs based on a minimum distance criterion. Moreover, the use of a least absolute selection and shrinkage operator (Lasso)-based variable selection algorithm in combination with the GPR model was presented for IFD \cite{Du2018Fault}. In \cite{Niu2019Fault}, PCA was utilized to detect IFs such as poor
contact faults in multi-axle speed sensors of high-speed trains. Note that in these methods, no time window was employed.

In \cite{Zhao2018Intermittent,Zhao2019Detecting}, the $T^2$ statistic and a generic quadratic-form statistic, combined with the moving average (MA) method, were utilized to detect IFs under the independent and multivariate Gaussian distribution assumption. In \cite{Li2016Fault}, dynamic PCA (DPCA) was applied to the measurement data of a gyroscope in order to detect its IFs. An exponentially weighted MA (EWMA)-based adaptive thresholding scheme was developed in \cite{Bakdi2017Anew} to detect IFs through $T^2$ and $Q$ statistics resulting from PCA. The adaptive threshold was updated by a modified EWMA control chart with limited window length, and was effective in reducing the fault clearance time delay between the real disappearance of IFs and the recovery of the fault indicator. The developed adaptive thresholding scheme was successfully applied to the detection of IFs in a cement rotary kiln \cite{Bakdi2017Fault}. In \cite{Bakdi2019Nonparametric}, a nonparametric Kullback-Leibler (KL) divergence resulting from multiblock PCA and moving window (MW)-based kernel density estimation (KDE) was presented to detect intermittent voltage sags in rooftop mounted PV (RMPV) systems. In \cite{Kammammettu2019Change}, the Kantorovich distance (KD), a metric originates from the optimal transport theory, between two sets of time series data (one of which employed an MW to update the online samples) was employed to detect IFs.

In practice, many IFs have small magnitudes and short durations \cite{Zhou2020Review}, which make them even more difficult to detect than incipient faults. Moreover, system dynamics and multi-level closed-loop control make industrial data autocorrelated. Due to the high-speed sampling requirement for capturing IFs, the property of non-independence in data is stronger and thus non-ignorable during IFD.
As a result, existing MSPM methods have the following problems that limit their application to IFD. On the one hand, static MSPM methods, which use only a single observation for FD such as the PCA-based control chart, have been found \cite{Shang2017Recursive,Lin2018Multimode,Kruger2007Improved} inefficient for small shifts, let alone IFs. Moreover, they cannot utilize autocorrelations in data.
On the other hand, dynamic MSPM methods such as DPCA and canonical variate analysis (CVA) consider a time sequence of measurements and can capture process dynamics (i.e, utilize autocorrelations). However, time lags are chosen only according to system orders, but not considering the characteristics of IFs (i.e., the fault duration and magnitude). Therefore, they may not gain enough sensitivity to IFs, and their efficiency of detecting intermittent small shifts still needs further study.

The MA-type scheme is a simple and powerful smoothing tool that can enhance the statistics' sensitivity to faults in practical applications, and is easy to integrate with many MSPM and machine learning methods. Two related schemes are the equally and exponentially weighted schemes. When samples are independent and identically distributed, the covariance matrices of the averaged sample after MA and EWMA are $1/W$ and $\lambda/(2-\lambda)$ of the covariance matrix of the original sample respectively \cite{Chen2001Principle,Ji2017Incipient}, where $W$ is the window length of MA and $0<\lambda\leq 1$ is the weighting factor of EWMA. This overall reduction of the covariance brings about the smoothing effects of MA-type schemes, and consequently improves the FD performance. However, this theoretical basis holds only for independent data, and the statistical basis for the use of MA-type schemes to smooth autocorrelated data is still lost.
In addition, both the MA and EWMA schemes have a fixed weighted form. So far, the weighted MA (WMA) scheme that allows putting different weights on samples within a time window for the purpose of FD has not been fully investigated. Moreover, the optimality of these weighted schemes in terms of fault detectability remains unproven. These issues constitute the main motivations of our study.

This paper investigates the IFD problem in weakly stationary processes. A time window and a weight vector are employed to increase the sensitivity to IFs, and the window length is selected considering the characteristics of IFs. Main contributions of the paper are summarized as follows:
1) An optimally weighted moving average $T^2$ control chart (OWMA-TCC) with stationary observations is proposed. Different from existing methods that put an equal weight on samples within a time window, OWMA-TCC uses correlation (autocorrelation and cross-correlation) information to find an optimal weight vector.
2) The concept of IF detectability is defined and corresponding detectability conditions are provided, which further serve as selection criteria of the optimal weight.
3) The optimal weight is given in the form of a solution to nonlinear equations, whose existence is proven with the help of the Brouwer fixed-point theory. Moreover, the uniqueness of the optimal weight is proven in several special cases.
4) We reveal that the optimal weight possesses a symmetrical structure, and the MA scheme is optimal for any IF directions when data are independent, which gives more explanations for the rationality of existing MA-based methods.
5) Comprehensive comparative studies with existing static and dynamic MSPM methods, such as PCA, MA-PCA, DPCA, CVA and MW-KD, are carried out on a numerical example and the benchmark continuous stirred tank reactor (CSTR) process, which illustrate the superior IFD performance of the OWMA-TCC.

The remainder of this paper is organized as follows. In Section \ref{WMATCCSec}, the WMA-TCC with stationary Gaussian observations is introduced for the IFD problem. Then, the detectability of IFs by the WMA-TCC is analyzed in Section \ref{DetectabilitySec}. The detectability conditions are further utilized to determine the optimal weight in Section \ref{OptimalWeightKnownXiSec}. Section \ref{NonGaussianSec} extends these results to weakly stationary processes without the Gaussianity assumption. Simulation results are presented in Section \ref{SimulationSec}, and conclusions are given in Section \ref{ConclusionSec}.

{\bf Notation:} Except where otherwise stated, the notations used throughout the paper are standard.
${\mathbb N_p}(\mu,\Sigma)$ represents a $p$-dimensional normal distribution with expectation $\mu$ and covariance matrix $\Sigma$.
${\mathbb W_p}(N,\Sigma)$ represents a $p$-dimensional Wishart distribution with $N$ degrees of freedom.
${\mathbb F}(p,N-p)$ is a central $F$ distribution with $p$ and $N-p$ degrees of freedom.
${\mathbb F}_{\alpha}(p,N-p)$ is the $1-\alpha$ percentile of the central $F$ distribution with $p$ and $N-p$ degrees of freedom.
${\mathbb {GP}}_p(\mu,R_l)$ represents a $p$-dimensional stationary Gaussian process with expectation $\mu$ and autocovariance function matrix $R_l$.
${\mathbb R}^{n}$ and ${\mathbb R}^{n \times m}$ denote the $n$-dimensional Euclidean space and the set of all $n \times m$ real matrices.
$\|\xi\|$ and $\|\xi\|_{\infty}$ denote the Euclidean norm and infinity norm of a vector $\xi$, respectively.
$A^T$, $A^{-1}$, $|A|$, ${\mathrm {tr}}(A)$ and \textrm{adj}($A$) stand for the transpose, the inverse, the determinant, the trace and the adjoint of a matrix $A$, respectively.
$\nabla_{\vec{a}_W}{\cal L}(\vec{a}_W,\lambda)$ is the gradient of ${\cal L}$ with respect to $\vec{a}_W$.
$\nabla^2_{\vec{a}_W}{\cal L}(\vec{a}_W,\lambda)$ is the Hessian matrix of ${\cal L}$ with respect to $\vec{a}_W$.
Scalars $a_1\cdots a_W$ form a row vector by $[a_1,a_2,\cdots,a_W]$, and form a column vector by $[a_1;a_2;\cdots;a_W]$. $\triangleq$ is to give definition.
$H_{l,l'}$ or $[H]_{l,l'}$ is an element of matrix $H$ located in the $l$th row and $l'$th column. $H_{l,:}$ and $H_{:,l}$ are the $l$th row and $l$th column of matrix $H$, respectively.
$\mathcal{T}_{\backslash i\backslash j}$ is the matrix obtained from $\mathcal{T}$ by deleting the row and column containing $\mathcal{T}_{i,j}$.
$I_{p}$ and $e_{pi}$ denote the $p$-dimensional identity matrix and its $i$th column, respectively; $1_W$ and $0_W$ denote the $W$-dimensional column vectors with all of its entries being one and zero, respectively.
The symbol $\otimes$ denotes the Kronecker product and $\delta_{ij}$ is the Kronecker function.
$\lambda_{\textrm{min}}(\Gamma)$ and $\lambda_{\textrm{max}}(\Gamma)$ are the minimum and maximum eigenvalues of matrix $\Gamma$, respectively.
$A\prec B$ and $A\preceq B$ mean that $A-B$ is negative definite and negative semidefinite, respectively.


\section{Methodology}\label{WMATCCSec}
When the process is under steady-state operation and no operators change the process dynamics, the acquired data tends to be stationary, non-anomalous, and with no trends \cite{Bakdi2017Fault,Bakdi2017Anew}. Thus, the dynamics of practical in-control systems can be approximated by a stationary stochastic process.
In this section, the WMA-TCC with stationary Gaussian observations is proposed for the purpose of FD in stationary Gaussian processes. The WMA-TCC in weakly stationary processes without the Gaussianity assumption is given in Section \ref{NonGaussianSec}.

\subsection{Preliminaries}\label{PreliminariesSub}
The following lemma is the key result regarding Hotelling's $T^2$ distribution, see \cite{Anderson2003An}.
\begin{lemma}\label{LemHotellingT}
Let $T^2=X^TS^{-1}X$, where $X$ and $S$ are independently distributed random variables with $X\sim{\mathbb N_p}(\mu,\Sigma)$ and $NS\sim{\mathbb W_p}(N,\Sigma)$, where $N\geq{p}$. Then
\begin{align}
T^2\sim \frac{Np}{N-p+1}{\mathbb F}(p,N-p+1;\epsilon^2),
\end{align}
where the noncentrality parameter $\epsilon^2=\mu^T\Sigma^{-1}\mu$.
\end{lemma}

\subsection{Weighted moving average $T^2$ control chart}\label{WMATCCSub}
The IFD task with stationary Gaussian observations concerns the analysis of latest $W$ new current process data $X^f_{k-W+1},\cdots,X^f_{k-1},X^f_k\in{\mathbb R}^{p}$ at each time $k$, to determine whether the process is statistically fault-free or not.
Different from existing MA-type schemes \cite{Chen2001Principle,Ji2016Incipient,Ji2017Incipient} that ordinarily have independence and identically Gaussian distribution assumptions, we here assume that systems' normal operation follows a stationary Gaussian process ${\mathbb {GP}}_p(\mu_f,R_l)$, whose autocovariance function reduces to nearly zero for large time lags. That is, for all $k$, ${\mathbb E}(X^f_k)=\mu_f$ and the autocovariance function ${\mathbb C\rm{ov}}(X^f_k,X^f_{k-l})=R_l$ depends only on the lag $l$. Moreover, we have $\|{R_l}\|\thickapprox 0$ for large $l$.

To construct the WMA-TCC, we collect $N$ sets of $W$ consecutive observations $X^j_i$, $i=1,2,\cdots,N$, $j=W,W\!-\!1,\cdots,1$ from the stationary Gaussian process ${\mathbb {GP}}_p(\mu,R_l)$ as training data, which can represent the statistic characteristics of systems' normal operating conditions. Moreover, $X^{j_1}_{i_1}$ and $X^{j_2}_{i_2}$ are independent and identically distributed for $i_1\!\neq\!i_2$. This can be achieved by taking samples with long enough intervals between different sets, and thus $\lim_{l\rightarrow\infty}\|{R_l}\|\!=\!0$. Note that in the same set, the sampling rate of training data should be equal to that of current process data. To sum up, the sampling strategy for training data is shown in \eqref{SampleTraining}, where $\widetilde{\cdots}$ means a long enough interval.
\begin{align}\label{SampleTraining}
\{X^W_1,X^{W-1}_1,\cdots,X^1_1\},\ \widetilde{\cdots}&\ ,\{X^W_2,X^{W-1}_2,\cdots,X^1_2\},\ \widetilde{\cdots}\nonumber\\
&\vdots\\
\{\underbrace{X^W_{N-1},X^{W-1}_{N-1},\cdots,X^1_{N-1}}\},\ &\widetilde{\cdots}\ ,\{\underbrace{X^W_N,X^{W-1}_N,\cdots,X^1_N}\}.\nonumber\\
\{a_W,a_{W-1},\cdots,a_1\}\quad\ \;&\quad\ \quad\{a_W,a_{W-1},\cdots,a_1\}\nonumber
\end{align}

The IFD problem can be viewed as a hypothesis testing problem concerning $H_0: \mu_f=\mu$ versus $H_1: \mu_f\neq\mu$.
Let $\vec{a}_W=[a_1,a_2,\cdots,a_W]^T$ be the weight vector. For the WMA-TCC, we put different weights on samples in the time window, as shown in \eqref{SampleTraining} and \eqref{SampleTest}.
\begin{align}\label{SampleTest}
\cdots,X^f_{k-W},&\{\underbrace{X^f_{k-W+1},X^f_{k-W+2},\cdots,X^f_k}\},X^f_{k+1},\cdots\nonumber\\
&\quad\ \{a_W,a_{W-1},\cdots,a_1\}
\end{align}

In practice, parameters $\mu_f, \mu, R_l$ are unknown, and we only know the sample means $\tilde{X}, \tilde{X}^f_k$ and the sample covariance matrix $\tilde{S}_W$ instead:
\begin{align}\label{WeightedSampleCov}
&\tilde{X}^f_k=\sum\limits_{j=1}^{W}a_jX^f_{k-j+1},\; \tilde{X}_i=\sum\limits_{j=1}^{W}a_jX^{j}_i,\; \tilde{X}=\frac{1}{N}\sum\limits_{i=1}^{N}\tilde{X}_i,\nonumber\\
&\tilde{S}_W=\frac{1}{N-1}\sum\limits_{i=1}^{N}(\tilde{X}_i-\tilde{X})(\tilde{X}_i-\tilde{X})^T,\ \sum\limits_{j=1}^{W}a_j=1.
\end{align}
Here, $\tilde{X}^f_k,\tilde{X}_i,\tilde{X},\tilde{S}_W$ are abbreviations for $\tilde{X}^f_k(\vec{a}_W)$, $\tilde{X}_i(\vec{a}_W)$, $\tilde{X}(\vec{a}_W)$, $\tilde{S}_W(\vec{a}_W)$ respectively, since they are actually matrix- or vector-valued functions of $\vec{a}_W$.
We also know that the sample means $\tilde{X}^f_k, \tilde{X}$ and the sample covariance matrix $\tilde{S}_W$ are independently distributed, with
\begin{align}\label{WeightedSampleDistribution}
(N-1)\tilde{S}_W\sim&{\mathbb W_p}(N-1,\tilde\Sigma_W),\ \tilde\Sigma_W\!=\!\sum\limits_{i=1}^{W}\sum\limits_{j=1}^{W}{a_i}{a_j}R_{i-j},\nonumber\\
(\tilde{X}^f_k-\tilde{X})&\sim{\mathbb N_p}(\mu_f-\mu,\frac{N+1}{N}\tilde\Sigma_W),
\end{align}
where $\tilde\Sigma_W$ is an abbreviation for $\tilde\Sigma_W(\vec{a}_W)$.

According to Lemma \ref{LemHotellingT}, the WMA-TCC with window length $W$, denoted as WMA-TCC($W$), with stationary Gaussian observations at time instance $k$ is then
\begin{align}\label{T2Conse}
\tilde{T}^2_k(W)&=(\tilde{X}^f_k-\tilde{X})^T\tilde{S}^{-1}_W(\tilde{X}^f_k-\tilde{X})\nonumber\\
&\sim\frac{p(N^2-1)}{N(N-p)}{\mathbb F}(p,N-p).
\end{align}
Here, we assume that $\tilde{S}_W(\vec{a}_W)$ is nonsingular for any weight vector $\vec{a}_W\neq 0_W$. Detailed explanations are given in \textit{Assumption \ref{AssmGammaW}} and \textit{Proposition \ref{ProSwHatTNonsingular}} of Section \ref{OptimalWeightKnownXiSec}.
For a given significance level $\alpha$, the process is considered normal at time instance $k$, i.e., to accept $H_0: \mu_f=\mu$, if
\begin{align}\label{UclConsecutive}
\tilde{T}^2_k(W)\leq\delta^2=\frac{p(N^2-1)}{N(N-p)}{\mathbb F}_{\alpha}(p,N-p),
\end{align}
where $\delta^2$ is the control limit of the WMA-TCC($W$). Otherwise, an alarm occurs at time instance $k$.
Inequality \eqref{UclConsecutive} gives the acceptance region of the hypothesis test. In Section \ref{NonGaussianSec}, the above WMA-TCC is generalized to weakly stationary processes without the Gaussianity assumption.

\section{Detectability analysis}\label{DetectabilitySec}
For the WMA-TCC, the window length and the weight vector are crucial parameters that can directly affect the IFD performance.
They should be carefully selected so that the detection capability for IFs is maximized. Thus, in this section, we analyze the IF detectability.

\subsection{Guaranteed detectability}
Consider the following widely used fault model in the MSPM framework \cite{Chen2018Deep,Alcala2009Reconstruction,Shang2018Recursive,Chen2019Data}:
\begin{align}\label{FaultModel}
X^f_k=X^{*}_k+\Xi_{k}F_k,
\end{align}
where $\Xi_{k}$ represents the fault direction, $\|F_k\|$ represents the fault magnitude, and $X^*_k$ represents the process fluctuation under normal conditions, all in time instance $k$. Note that the above fault model can represent a multiple fault when the rank of the column vector $F_k$ is larger than one.
By introducing the time window, we have
\begin{align}\label{FaultModelWin}
\tilde{X}^f_k=\tilde{X}^{*}_k+\tilde{\Xi}_{k}\tilde{F}_k,\; \tilde{X}^*_k&=\sum\limits_{j=1}^{W}a_jX^*_{k-j+1},
\end{align}
where $\tilde\Xi_k\tilde{F}_k$ is the effect of all faults in the time window, and $\tilde{X}^*_k\sim{\mathbb N_p}(\mu,\tilde\Sigma_W)$.
When we analyze the fault detectability, the following condition is introduced:
\begin{align}\label{AssuAcceptRegion}
\|\tilde{S}^{-1/2}_W(\tilde{X}^{*}_{k}-\tilde{X})\|^2\leq\delta^2.
\end{align}
\begin{remark}
Inequality \eqref{AssuAcceptRegion} is commonly employed by literature addressing fault detectability problems in the MSPM framework \cite{Qin2003Statistical,Alcala2009Reconstruction,Mnassri2013Generalization,Ji2017Incipient}.
The condition means that the fault-free process $\tilde{X}^*_k$ fluctuates within its acceptance region \eqref{UclConsecutive}. Since a small significance level (i.e., $\alpha=0.01$) is always selected, this condition holds with high probability. Note that this condition is only introduced to analyze detectability, and thus has no limitation to the practical application of the method.
\end{remark}

\begin{figure}
\begin{center}
\includegraphics[width=8.4cm]{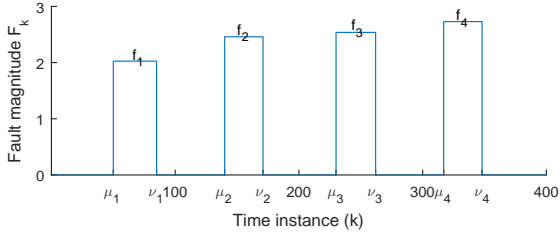}    
\caption{An example of intermittent faults.}
\label{IFexample}
\end{center}
\end{figure}

Since $\Xi_{k}F_k\in{\mathbb R}^{p}$, it can be denoted by a scalar and a column vector whose norm is one. Then, in the case of IFs, as shown in Fig.~\ref{IFexample}, the corresponding fault model can be represented \cite{Isermann2005Model,Zhang2020Intermittent,Zhou2020Review} by
\begin{align}\label{IFModel}
\Xi_{k}F_k=\sum\limits_{q=1}^{\infty}\left[\Gamma(k-\mu_{q})-\Gamma(k-\nu_{q})\right]\xi_q{f_q},
\end{align}
where $\Gamma(\cdot)$ is the step function; $\mu_{q}$, $\nu_{q}$ represent the appearing and disappearing time of the $q$th IF, satisfying $\mu_q<\nu_q<\mu_{q+1}$; and $\xi_q\in{\mathbb R}^{p}$, $f_q\in{\mathbb R}^{1}$ are the direction and magnitude of the $q$th IF, satisfying $\|\xi_q\|\!=\!1$.
Moreover, the active and inactive duration of the $q$th IF are $\tau^o_q=\nu_q-\mu_q$ and $\tau^r_q=\mu_{q+1}-\nu_q$, respectively. Note that they are counted by sampling intervals here. The $q$th IF can be denoted by five parameters, i.e., IF$(\xi_q, f_q, \tau^r_{q-1}, \tau^o_q, \tau^r_q)$.
\begin{remark}
Recall that the characteristics of IFs are small magnitude and short duration. In most cases, since the fault magnitude is small, when an IF becomes active, after exhibiting a short transient behavior, the system will be driven to another steady state soon by the closed-loop control, instead of being continuously sharp fluctuations or out of control. Similarly, when the IF becomes inactive, after a short transition, the closed-loop control will drive the system back to its normal steady state soon. Moreover, since the fault duration is short, we can assume the fault direction and magnitude within each IF to be constant. Therefore, IFs can be represented by the form of intermittent biases as \eqref{IFModel}. This statement will be confirmed by a realistic simulation of the practical CSTR benchmark in Section \ref{SimulationSec}.
\end{remark}

The fault detectability concept was first defined in \cite{Dunia1998A,Dunia1998Subspace} within the MSPM framework, and has been widely adopted by a variety of MSPM methods \cite{Qin2003Statistical,Alcala2009Reconstruction,Mnassri2013Generalization,Ji2017Incipient} to study the FD performance. However, the concept has been mainly concerned with PFs.
Compared with a PFD task, additional requirements for an IFD task \cite{Biswas2012diagnosability,Zhang2019Robust,Zhou2020Review} are to determine each appearance (disappearance) of an IF before its subsequent disappearance (appearance), otherwise missing or false alarms occur.
Following these considerations, this paper extends and generalizes the original fault detectability concept \cite{Dunia1998A} to make it suitable for both PFs and IFs.

\begin{definition}\label{DefnIFDrec}
For a given significance level $\alpha$, the disappearance of the $q$th IF is said to be \textbf{guaranteed detectable} (DPG-detectable) by the WMA-TCC($W$), if there exists a time instance $\nu_q\!\leq\!\bm{k}^{\#}\!<\!\mu_{q+1}$ such that for each $k^{\#}\leq\bm{k}<\mu_{q+1}$, the detection statistic $\tilde{T}^2_k(W)\leq\delta^2$ is guaranteed for all values of $\tilde{X}^*_k$ in \eqref{AssuAcceptRegion}.
\end{definition}

\begin{definition}\label{DefnIFDocc}
For a given significance level $\alpha$, the appearance of the $q$th IF is said to be \textbf{guaranteed detectable} (APG-detectable) by the WMA-TCC($W$), if the disappearance of the $(q\!-\!1)$th IF is guaranteed detectable, and there exists a time instance $\mu_{q}\!\leq\bm{k}^*\!<\!\nu_{q}$ such that for each ${k^*}\leq\bm{k}<\nu_{q}$, the detection statistic $\tilde{T}^2_k(W)>\delta^2$ is guaranteed for all values of $\tilde{X}^*_k$ in \eqref{AssuAcceptRegion}.
\end{definition}

\begin{definition}\label{DefnIFD}
For a given significance level $\alpha$, the $q$th IF is said to be \textbf{guaranteed detectable} (G-detectable) by the WMA-TCC($W$), if both the appearance and disappearance of the $q$th IF are guaranteed detectable.
\end{definition}

\subsection{Detectability conditions}
Intuitively, to detect the disappearance/appearance of an IF, we can choose a window length that is no more than the IF's inactive/active duration, so that the WMA-TCC($W$) is free from interference of previous faulty/fault-free samples after some delay.

\begin{lemma}\label{LemIFDrec}
For the WMA-TCC($W$) and a given significance level $\alpha$, when $W\leq\tau^r_{q}$, the disappearance of the $q$th IF is guaranteed detectable {(DPG-detectable)}.
\end{lemma}
\textbf{Proof.} According to the IF model (\ref{IFModel}), when $W\leq\tau^r_{q}$, there exists a time instance $\nu_q\leq\bm{k}^{\#}<\mu_{q+1}$, such that for each $k^{\#}\leq\bm{k}<\mu_{q+1}$, all $W$ current process samples within the time window are fault-free. Then we have $\tilde{X}^f_k=\tilde{X}^{*}_{k}$ and
\begin{align*}
\tilde{T}^2_k(W)=\|\tilde{S}^{-1/2}_W(\tilde{X}^f_k-\tilde{X})\|^2=\|\tilde{S}^{-1/2}_W(\tilde{X}^{*}_{k}-\tilde{X})\|^2.
\end{align*}
Thus, for each $k^{\#}\leq\bm{k}<\mu_{q+1}$, the detection statistic $\tilde{T}^2_k(W)\leq\delta^2$ is guaranteed for all values of $\tilde{X}^*_k$ in \eqref{AssuAcceptRegion}. \qed

\begin{lemma}\label{LemIFDocc}
For the WMA-TCC($W$) and a given significance level $\alpha$, when $W\leq\min\{\tau^r_{q-1},\tau^o_{q}\}$, the appearance of the $q$th IF is guaranteed detectable {(APG-detectable)} if and only if
\begin{align}\label{EquIFDocc}
\left\|{\tilde{S}_W^{-1/2}\xi_q{f_q}}\right\|>2\delta.
\end{align}
\end{lemma}
\textbf{Proof.} According to Lemma \ref{LemIFDrec}, when $W\!\leq\!\min\{\tau^r_{q-1},\tau^o_{q}\}$, the disappearance of the $(q\!-\!1)$th IF is guaranteed detectable. Moreover, there exists a time instance $\mu_{q}\!\leq\bm{k^*}\!<\!\nu_{q}$, such that for each $k^*\!\leq\!\bm{k}\!<\!\nu_{q}$, all $W$ current process samples within the time window are faulty. Then we have $\tilde{X}^f_k=\tilde{X}^{*}_{k}+\xi_q{f_q}$ and
\begin{align}\label{T2LowerIF}
\tilde{T}^2_k(W)&=\|\tilde{S}^{-1/2}_W(\tilde{X}^{*}_{k}-\tilde{X}+\xi_q{f_q})\|^2 \nonumber\\
&\geq\left( \|\tilde{S}_W^{-1/2}\xi_q{f_q}\|-\|\tilde{S}^{-1/2}_W(\tilde{X}^{*}_{k}-\tilde{X})\| \right)^2.
\end{align}
Then by following \eqref{EquIFDocc}, \eqref{AssuAcceptRegion} and \eqref{T2LowerIF}, we derive that for each ${k^*}\!\leq\!\bm{k}<\nu_{q}$, $\tilde{T}^2_k(W)\!>\!\delta^2$ is guaranteed for all values of $\tilde{X}^*_k$ in \eqref{AssuAcceptRegion} and the proof of sufficiency is complete.

We now prove the necessity by contraposition. The contrapositive of the necessity statement is:
\textit{When $W\!\leq\!\min\{\tau^r_{q-1},\tau^o_{q}\}$, if $\left\|{\tilde{S}_W^{-1/2}\xi_q{f_q}}\right\|\!\leq\!2\delta$, then the disappearance of the $(q\!-\!1)$th IF is not guaranteed detectable, or for any time instance $\mu_{q}\!\leq\bm{k^*}\!<\!\nu_{q}$, there exists a time instance $k^*\!\leq\!\bm{k}\!<\!\nu_{q}$ and a value of $\tilde{X}^*_k$ in \eqref{AssuAcceptRegion}, making $\tilde{T}^2_k(W)\leq\delta^2$ valid.}
This contrapositive statement can be proven as follows.
For any given $\mu_{q}\!\leq\bm{k^*}\!<\!\nu_{q}$, we consider time instance $k\!=\!\nu_{q}\!-\!1$ which satisfies $k^*\!\leq\!\bm{k}\!<\!\nu_{q}$.
We further consider the following value of $\tilde{X}^*_k$: $\tilde{S}^{-1/2}_W(\tilde{X}^{*}_{k}-\tilde{X})=-\tilde{S}_W^{-1/2}\xi_q{f_q}/2$, which satisfies \eqref{AssuAcceptRegion} if $\|\tilde{S}_W^{-1/2}\xi_q{f_q}\|\!\leq\!2\delta$.
Note that at time instance $k\!=\!\nu_{q}\!-\!1$, we have $\tilde{X}^f_k=\tilde{X}^{*}_{k}+\xi_q{f_q}$ and consequently
$\tilde{T}^2_k(W)=\|\tilde{S}_W^{-1/2}\xi_q{f_q}/2\|^2\leq\delta^2$. Having proven the contrapositive, we infer the original statement and the proof of necessity is complete. \qed

\begin{theorem}\label{ThmIFD0}
For the WMA-TCC($W$) and a given significance level $\alpha$, when $W\!\leq\!W^\#\!\triangleq\!\min\{\tau^r_{q-1},\tau^o_{q},\tau^r_{q}\}$, the $q$th IF is guaranteed detectable {(G-detectable)} if and only if inequality (\ref{EquIFDocc}) holds.
\end{theorem}
\textbf{Proof.} Directly derived from Lemmas \ref{LemIFDrec} and \ref{LemIFDocc}. \qed

\section{Determination of the weight and window length}\label{OptimalWeightKnownXiSec}
In this section, methods to determine the weight vector and window length are provided, along with discussions on the existence, symmetry and uniqueness of the optimal weight.
\subsection{Problem formulation and main results}
Now, we are in the position to find the optimal weight vector based on the above derived detectability conditions, and present the main problem as follows.

\begin{problem}\label{PlmOptimalWeight}
For the WMA-TCC($W$), $W\leq W^\#$, find the optimal weight $\vec{a}^*_W$ that
\begin{align}
\label{EquOptFunc}
\max_{\vec{a}_W}\quad&\beta(\vec{a}_W)=\frac{1}{2}\|\tilde{S}_W^{-1/2}\xi_q\|^2,\\
\label{EquWeightSumConstraint}
{\rm s.t.}\quad&g(\vec{a}_W)=\sum\limits_{j=1}^{W}a_j=1.
\end{align}
\end{problem}

\begin{theorem}\label{ThmOptimalWeight}
The optimal weight $\vec{a}^*_W$ maximizing $\beta(\vec{a}_W)$ of Problem \ref{PlmOptimalWeight} satisfies
\begin{align}\label{EquOptimalWeightFstNesCond}
\hat{\mathcal{T}}(\vec{a}^*_W)\vec{a}^*_W=b,
\end{align}
and
\begin{align}\label{EquOptimalWeightSecNesCond}
(-1)^{k}\left| \overline{{\cal H}}_{k}(\vec{a}^*_W) \right|\geq 0,\quad k=2,3,\cdots,W,
\end{align}
where $\hat{\mathcal{T}}(\vec{a}_W)\in{\mathbb R}^{W\times W}, b=[0,\cdots,0,1]^T\in{\mathbb R}^{W}$,
\begin{align}\label{EquOptimalWeightFstNesCondSpt}
\hat{\mathcal{T}}_{l,j}(\vec{a}_W)&=\left\{\begin{array}{ll}
\xi^T_q\tilde{S}_W^{-1}\left(\hat{R}_{lj}-\hat{R}_{(l+1)j}\right)\tilde{S}_W^{-1}\xi_q,&{}l<W,\\
1,&{}l=W,
\end{array}\right.\nonumber\\
\hat{R}_{lj}&=\frac{1}{N-1}\sum\limits_{i=1}^{N}(X^l_i-\bar{X}^l)(X^j_i-\bar{X}^j)^T,\nonumber\\
\bar{X}^j&=\frac{1}{N}\sum\limits_{i=1}^{N}X^j_i,
\end{align}
and $\tilde{S}_W$ is short for $\tilde{S}_W(\vec{a}_W)$ calculated by \eqref{WeightedSampleCov},
\begin{align}\label{EquOptimalWeightSecNesCondSpt}
\left|\overline{{\cal H}}_{k}(\vec{a}_W) \right|&=
\left| {\begin{array}{cccc}
0 & 1 & \cdots & 1\\
1 & \hat{H}_{1,1}(\vec{a}_W) & \cdots & \hat{H}_{1,k}(\vec{a}_W)\\
\vdots & \vdots & \ddots & \vdots\\
1 & \hat{H}_{k,1}(\vec{a}_W) & \cdots & \hat{H}_{k,k}(\vec{a}_W)
\end{array}} \right|,\nonumber\\
\hat{H}_{l,l'}(\vec{a}_W)&=h^T_lh_{l'}-\xi^T_q\tilde{S}_W^{-1}\hat{R}_{ll'}\tilde{S}_W^{-1}\xi_q,\nonumber\\
\hat{h}_l(\vec{a}_W)&=\tilde{S}_W^{-1/2}\left(\sum\limits_{j=1}^{W}a_j(\hat{R}_{lj}+\hat{R}^T_{lj})\right)\tilde{S}_W^{-1}\xi_q.
\end{align}
\end{theorem}
\textbf{Proof.} For this nonlinear constrained optimization problem, we can construct a Lagrange function given by
\begin{align}
{\cal L}(\vec{a}_W,\lambda)=\frac{1}{2}\|\tilde{S}_W^{-1/2}\xi_q\|^2+\lambda(\sum\limits_{j=1}^{W}a_j-1),
\end{align}
where $\lambda$ is a Lagrange multiplier. According to the Karush-Kuhn-Tucker conditions (first-order necessary conditions) \cite{Luenberger2008Linear}, the optimal weight $\vec{a}^*_W$ should satisfy
\begin{align}
\nabla_{\vec{a}_W}{\cal L}(\vec{a}_W,\lambda)=0_W, \quad \nabla_{\lambda}{\cal L}(\vec{a}_W,\lambda)=0.
\end{align}
Note that
\begin{align}
\frac{\partial{\cal L}(\vec{a}_W,\lambda)}{\partial a_l}
&=-\frac{1}{2}\xi^T_q\tilde{S}_W^{-1}\left(\frac{\partial\tilde{S}_W}{\partial a_l}\right)\tilde{S}_W^{-1}\xi_q+\lambda\nonumber\\
&=-\xi^T_q\tilde{S}_W^{-1}\left(\sum\limits_{j=1}^{W}a_j\hat{R}_{lj}\right)\tilde{S}_W^{-1}\xi_q+\lambda.
\end{align}
By setting the above derivative of ${\cal L}(\vec{a}_W,\lambda)$ with respect to $\vec{a}_W$ to zeros, the following equations can be obtained.
\begin{align}
\label{DerivativeEqual}
\xi^T_q\tilde{S}_W^{-1}\!\left(\sum\limits_{j=1}^{W}a_j(\hat{R}_{lj}-\hat{R}_{l'j})\right)\!\tilde{S}_W^{-1}\xi_q\!=\!0,\ 1\leq l,l'\leq W.
\end{align}
Thus, integrating \eqref{DerivativeEqual} with \eqref{EquWeightSumConstraint}, the first-order necessary conditions for the constrained optimization problem are derived as \eqref{EquOptimalWeightFstNesCond}.

When $\vec{a}^*_W$ meets \eqref{EquOptimalWeightFstNesCond}, it is considered an extremum point or saddle point for function \eqref{EquOptFunc} subject to constraint \eqref{EquWeightSumConstraint}.
According to \cite{Chiang2005Fundamental}, second-order necessary conditions for $\vec{a}^*_W$ to be a maximum point are: the leading principal minors of $\overline{{\cal H}}(\vec{a}^*_W)$ of order $k+\!1$ ($k\!=\!2,3,\!\cdots\!,\!W$) have sign $(-1)^{k}$ or equal to zero, where
\begin{align}\label{BorderedHessianMatrix}
\overline{{\cal H}}(\vec{a}_W)=
\left[ {\begin{array}{cc}
0 & \nabla^T_{\vec{a}_W}{g}(\vec{a}_W)\\
\nabla_{\vec{a}_W}{g}(\vec{a}_W) & \hat{H}(\vec{a}_W)
\end{array}} \right],
\end{align}
is a bordered Hessian matrix and
\begin{align*}
\hat{H}(\vec{a}_W)\!=\!\nabla^2_{\vec{a}_W}{\cal L}(\vec{a}_W,\lambda),\;\textrm{i.e.}\;
\hat{H}_{l,l'}(\vec{a}_W)\!=\!\frac{\partial^2{\cal L}(\vec{a}_W,\lambda)}{\partial a_l \partial a_{l'}}.
\end{align*}
Thus, the second-order necessary conditions for the optimization problem are derived as \eqref{EquOptimalWeightSecNesCond}. \qed

\subsection{Existence of the solution}
In this subsection, we prove the existence of the solution of nonlinear equations \eqref{EquOptimalWeightFstNesCond} with the help of the well-known Brouwer fixed-point theory.
We begin with the following assumption and the result is given in \textit{Theorem \ref{ThmSolutionExistNnequ}} at last. Additionally, methods to obtain the optimal weight are discussed and a bound of the optimal weight is given.
\begin{assumption}\label{AssmGammaW}
$\hat{\Gamma}^W$ is nonsingular, where
\begin{align}\label{HatGammak}
\hat{\Gamma}^k&=\left[ {\begin{array}{cccc}
\hat{R}_{11} & \hat{R}_{12} & \cdots & \hat{R}_{1k}\\
\hat{R}_{21} & \hat{R}_{22} & \cdots & \hat{R}_{2k}\\
\vdots & \vdots & \ddots & \vdots \\
\hat{R}_{k1} & \hat{R}_{k2} & \cdots & \hat{R}_{kk}\end{array}} \right]\in{\mathbb R}^{pk\times pk}.
\end{align}
\end{assumption}
\begin{remark}
\textit{Assumption \ref{AssmGammaW}} is the same as the assumption for Yule-Walker equations, which are well-known in the field of parameter identification of time series models. In real applications, \textit{Assumption \ref{AssmGammaW}} holds due to the existence of process and measurement noises.
\end{remark}

\begin{proposition}\label{ProSwHatTNonsingular}
Suppose \textit{Assumption \ref{AssmGammaW}} holds, then $\tilde{S}_W(\vec{a}_W)$ and $\hat{\mathcal{T}}(\vec{a}_W)$ are nonsingular, if $\vec{a}_W\neq 0_W$ and $\|\vec{a}_W\|<\infty$.
\end{proposition}
\textbf{Proof.} Let $\varepsilon^j_i=X^j_i-\bar{X}^j$. Then, we can rewrite $\hat{\Gamma}^k=\frac{1}{N-1}\hat{\Upsilon}_k\hat{\Upsilon}^T_k$, where $\hat{\Upsilon}_k\in{\mathbb R}^{pk\times N}$ and
\begin{align*}
\hat{\Upsilon}_k=\left[ {\begin{array}{cccc}
\varepsilon^1_1 & \varepsilon^1_2 & \cdots & \varepsilon^1_N\\
\varepsilon^2_1 & \varepsilon^2_2 & \cdots & \varepsilon^2_N\\
\vdots & \vdots & \ddots & \vdots\\
\varepsilon^k_1 & \varepsilon^k_2 & \cdots & \varepsilon^k_N \end{array}} \right].
\end{align*}
Thus, $\hat{\Gamma}^k$ is positive semidefinite. Moreover, by following \textit{Assumption \ref{AssmGammaW}}, we know that $\hat{\Gamma}^W$ is positive definite. According to \eqref{WeightedSampleCov} and \eqref{EquOptimalWeightFstNesCondSpt}, we have
\begin{align}\label{SwHatR}
&\tilde{S}_W(\vec{a}_W)=\frac{1}{N-1}\sum\limits_{k=1}^{N}(\tilde{X}_k-\tilde{X})(\tilde{X}_k-\tilde{X})^T\nonumber\\
&=\!\frac{1}{N\!-\!1}\!\sum\limits_{k=1}^{N}\left[\sum\limits_{i=1}^{W}\!a_i(X^i_k\!-\!\frac{1}{N}\!\sum\limits_{l=1}^{N}\!X^i_l)\right]\left[\sum\limits_{j=1}^{W}\!a_j(X^j_k\!-\!\frac{1}{N}\!\sum\limits_{l=1}^{N}\!X^j_l)\right]^T\nonumber\\
&=\frac{1}{N-1}\sum\limits_{k=1}^{N}\sum\limits_{i=1}^{W}\sum\limits_{j=1}^{W}a_ia_j(X^i_k-\bar{X}^i)(X^j_k-\bar{X}^j)^T\nonumber\\
&=\sum\limits_{i=1}^{W}\sum\limits_{j=1}^{W}{a_i}{a_j}\hat{R}_{ij}=\left(\vec{a}_W\otimes I_p\right)^T\hat{\Gamma}^W\left(\vec{a}_W\otimes I_p\right).
\end{align}
For any $\vec{a}_W\neq 0_W$, the matrix $\vec{a}_W\otimes I_p$ is full column rank. Thus, $\tilde{S}_W(\vec{a}_W)$ is nonsingular, and
\begin{align}\label{SwBound}
0<\lambda_{\min}(\hat{\Gamma}^W)\|\vec{a}_W\|^2I_{p}\leq\tilde{S}_W\leq\lambda_{\max}(\hat{\Gamma}^W)\|\vec{a}_W\|^2I_{p}.
\end{align}
Let $\hat{\gamma}^{W}\in{\mathbb R}^{W\times W}$ be the abbreviation of $\hat{\gamma}^{W}(\vec{a}_W)$, and define
\begin{align}\label{gammaHatR}
\hat{\gamma}^{W}_{l,j}=\xi^T_q\tilde{S}_W^{-1}\hat{R}_{lj}\tilde{S}_W^{-1}\xi_q.
\end{align}
Then, it follows from \textit{Assumption \ref{AssmGammaW}} and \eqref{SwBound} that, for any $\vec{a}_W\neq 0_W$ and $\|\vec{a}_W\|<\infty$,
\begin{align*}
\hat{\gamma}^{W}=\left(I_W\otimes\tilde{S}^{-1}_W\xi_q\right)^T\hat{\Gamma}^W\left(I_W\otimes\tilde{S}^{-1}_W\xi_q\right),
\end{align*}
is nonsingular and positive definite.
By following a few reformulations, we can rewrite $\hat{\mathcal{T}}(\vec{a}_W)=\hat{J}\hat{\gamma}^{W}$, where
\begin{align*}
\hat{J}=\left[ \begin{array}{c}
\underline{\begin{array}{ccccc}
1 & -1 & 0 & \cdots & 0\\
0 & 1 & -1 & \ddots & \vdots\\
\vdots & \ddots & \ddots & \ddots & 0\\
0 & \cdots & 0 & 1 & -1\end{array}} \\
1^T_W(\hat{\gamma}^{W})^{-1}
\end{array}\right]\in{\mathbb R}^{W\times W}.
\end{align*}
Thus, $\hat{\mathcal{T}}(\vec{a}_W)$ is nonsingular if and only if $\hat{J}$ is nonsingular.
We assume that $\hat{J}$ is singular, then there exists $\vec{\alpha}_W=[\alpha_1,\alpha_2,\cdots,\alpha_W]^T\neq 0_W$, such that
\begin{align*}
\alpha_1\hat{J}_{1,:}+\alpha_2\hat{J}_{2,:}+\cdots+\alpha_{W-1}\hat{J}_{W-1,:}+\alpha_W\hat{J}_{W,:}=0^T_W.
\end{align*}
Multiplying both sides by $1_W$ on the right, we have $\alpha_W1^T_W(\hat{\gamma}^{W})^{-1}1_W\!=\!0$. Since $\hat{\gamma}^{W}$ is positive definite, we obtain $\alpha_W=0$. This means that the first $W-1$ rows of $\hat{J}$ are linearly dependent, which contradicts the fact that $\hat{J}_{\backslash W\backslash\emptyset}$ has full row rank. Thus, $\hat{J}$ is nonsingular and the proof is complete. \qed

\begin{remark} According to \textit{Proposition \ref{ProSwHatTNonsingular}}, we can rewrite
\begin{align}\label{AwFixedPoint}
\vec{a}^*_W=\hat{\mathcal{T}}^{-1}(\vec{a}^*_W)b\triangleq\hat{\mathcal{F}}(\vec{a}^*_W)\in{\mathbb R}^{W}.
\end{align}
It can be seen that $\vec{a}^*_W$ is a fixed-point of function $\hat{\mathcal{F}}$. According to our practical experience, $\vec{a}^*_W$ can be obtained by successive approximations as follows
\begin{align}\label{AwIteration}
\vec{a}^{k+1}_W=\hat{\mathcal{F}}(\vec{a}^k_W),\ \forall\vec{a}^0_W\neq 0_W, \|\vec{a}^0_W\|<\infty.
\end{align}
Since $\hat{\mathcal{F}}(\vec{a}^k_W)\neq 0_W$ and $\|\hat{\mathcal{F}}(\vec{a}^k_W)\|<\infty$, \textit{Propositions \ref{ProSwHatTNonsingular}} and \ref{ProAwBounded} guarantee this process is always implementable.
Moreover, it follows from \eqref{SwBound} that $\lim_{\|\vec{a}_W\|\to\infty}\beta(\vec{a}_W)\!=\!0$. Thus, although any $\vec{a}_W$ such that $\|\vec{a}_W\|\!=\!\infty$ and $g(\vec{a}_W)=1$ satisfies \eqref{EquOptimalWeightFstNesCond}, it is not the solution of Problem \ref{PlmOptimalWeight}.
\end{remark}

\begin{lemma}\label{LemXAYIneq}
For any column vectors $x,y$ and matrix $P\succeq 0$, the following inequality holds:
\begin{align}
2\|x^TPy\|\leq x^TPx+y^TPy.
\end{align}
\end{lemma}
\textbf{Proof.} Directly derived from $0\leq (x-y)^TP(x-y)$ and $0\leq (x+y)^TP(x+y)$. \qed

\begin{proposition}\label{ProAwBounded}
Suppose \textit{Assumption \ref{AssmGammaW}} holds, then $\|\hat{\mathcal{F}}(\vec{a}_W)\|_{\infty}\leq\frac{W+1}{2W}\frac{\lambda_{\max}(\hat{\Gamma}^W)}{\lambda_{\min}(\hat{\Gamma}^W)}\triangleq d_W$ and $g\left( \hat{\mathcal{F}}(\vec{a}_W) \right)=1$, if $\vec{a}_W\neq 0_W$ and $\|\vec{a}_W\|<\infty$.
\end{proposition}
\textbf{Proof.} Let $\vec{c}_W=[c_1,\cdots,c_W]^T$, according to \eqref{AwFixedPoint}, we have
\begin{align}\label{hatFcW}
\hat{\mathcal{F}}(\vec{a}_W)\!=\!|\hat{\mathcal{T}}(\vec{a}_W)|^{-1}\textrm{adj}\left( \hat{\mathcal{T}}(\vec{a}_W) \right)b
\!=\!|\hat{\mathcal{T}}(\vec{a}_W)|^{-1}\vec{c}_W,
\end{align}
with $c_i\!=\!(-1)^{W+i}|\hat{\mathcal{T}}_{\backslash W\backslash i}(\vec{a}_W)|\!=\!|\check{\mathcal{T}}(\vec{a}_W,i)|$, where
\begin{align}\label{checkTi}
\check{\mathcal{T}}_{l,j}(\vec{a}_W,i)=\left\{\begin{array}{ll}
\hat{\mathcal{T}}_{l,j}(\vec{a}_W),&{}\quad l<W,\vspace{0.1cm}\\
\delta_{ij},&{}\quad l=W.
\end{array}\right.
\end{align}
Note that $g\left( \vec{c}_W \right)=|\hat{\mathcal{T}}(\vec{a}_W)|$. Thus, $g\left( \hat{\mathcal{F}}(\vec{a}_W) \right)=|\hat{\mathcal{T}}(\vec{a}_W)|^{-1}g\left( \vec{c}_W \right)=1$.
Moreover, by following a few reformulations, we can rewrite $\check{\mathcal{T}}(\vec{a}_W,i)=\check{J}^i\hat{\gamma}^{W}$, where
\begin{align*}
\check{J}^i=\left[ \begin{array}{c}
\underline{\begin{array}{ccccc}
1 & -1 & 0 & \cdots & 0\\
0 & 1 & -1 & \ddots & \vdots\\
\vdots & \ddots & \ddots & \ddots & 0\\
0 & \cdots & 0 & 1 & -1\end{array}} \\
e^T_{Wi}(\hat{\gamma}^{W})^{-1}
\end{array}\right]\in{\mathbb R}^{W\times W}.
\end{align*}
Note that $0<\lambda_{\min}(\hat{\Gamma}^W)I_{pW}\leq\hat{\Gamma}^W\leq\lambda_{\max}(\hat{\Gamma}^W)I_{pW}$. Then
\begin{align*}
0<\lambda_{\min}(\hat{\Gamma}^W)\varrho(\vec{a}_W)I_{W}\leq\hat{\gamma}^{W}\leq\lambda_{\max}(\hat{\Gamma}^W)\varrho(\vec{a}_W)I_{W},
\end{align*}
where $\varrho(\vec{a}_W)=\xi^T_q\tilde{S}^{-2}_W\xi_q$. For $|\hat{J}|$ and $|\check{J}^i|$, adding its $j$th column to its $j\!-\!1$th column in turn, we obtain $|\hat{J}|=1^T_W(\hat{\gamma}^{W})^{-1}1_W$ and $|\check{J}^i|=e^T_{Wi}(\hat{\gamma}^{W})^{-1}1_W$. Note that
\begin{align*}
\frac{W}{\lambda_{\max}(\hat{\Gamma}^W)\varrho(\vec{a}_W)}&\leq|\hat{J}|=1^T_W(\hat{\gamma}^{W})^{-1}1_W\leq \frac{W}{\lambda_{\min}(\hat{\Gamma}^W)\varrho(\vec{a}_W)},\\
\frac{1}{\lambda_{\max}(\hat{\Gamma}^W)\varrho(\vec{a}_W)}&\leq e^T_{Wi}(\hat{\gamma}^{W})^{-1}e_{Wi}\leq \frac{1}{\lambda_{\min}(\hat{\Gamma}^W)\varrho(\vec{a}_W)}.
\end{align*}
Then, according to \textit{Lemma \ref{LemXAYIneq}}, we have
\begin{align*}
\|c_i\|&=\left\||\check{J}^i|\right\||\hat{\gamma}^{W}|\leq\frac{1}{2}\left( e^T_{Wi}(\hat{\gamma}^{W})^{-1}e_{Wi}+|\hat{J}| \right)|\hat{\gamma}^{W}|\nonumber\\
&\leq \frac{W+1}{2\lambda_{\min}(\hat{\Gamma}^W)\varrho(\vec{a}_W)}|\hat{\gamma}^{W}|.
\end{align*}
For the $i$th element of $\hat{\mathcal{F}}(\vec{a}_W)$, we have
\begin{align*}
\|\hat{\mathcal{F}}_i(\vec{a}_W)\|=\frac{\|c_i\|}{|\hat{J}||\hat{\gamma}^{W}|}\leq\frac{W+1}{2W}\frac{\lambda_{\max}(\hat{\Gamma}^W)}{\lambda_{\min}(\hat{\Gamma}^W)}.
\end{align*}
Note that $\|\hat{\mathcal{F}}(\vec{a}_W)\|_{\infty}=\max_{i=1,\cdots,W} \|\hat{\mathcal{F}}_i(\vec{a}_W)\|$, then the proof is complete. \qed

\begin{remark} \textit{Proposition \ref{ProAwBounded}} presents a bound of the optimal weight, i.e., $\vec{a}^*_W\in\mathcal{M}_W$ given in \eqref{EquSolutionBound}. Moreover, \textit{Proposition \ref{ProAwBounded}} further guarantees the iteration process \eqref{AwIteration} is always bounded. In the following, we give the well-known Brouwer fixed-point theorem.
\end{remark}

\begin{lemma}\cite{Zeidler1986Nonlinear}
\label{LemFixedPointBrouwer}
Suppose that $M$ is a nonempty, convex, compact subset of ${\mathbb R}^{n}$, where $n\geq 1$, and that $\mathcal{F}: M\rightarrow M$ is a continuous mapping. Then $\mathcal{F}$ has a fixed point.
\end{lemma}

\begin{theorem}\label{ThmSolutionExistNnequ}
Suppose \textit{Assumption \ref{AssmGammaW}} holds, then the nonlinear equations \eqref{EquOptimalWeightFstNesCond} have a solution.
\end{theorem}
\textbf{Proof.} Define a subset of ${\mathbb R}^{W}$ as
\begin{align}\label{EquSolutionBound}
\mathcal{M}_W=\{\vec{a}_W\in{\mathbb R}^{W}: g(\vec{a}_W)=1,\ \|\vec{a}_W\|_{\infty}\leq d_W\},
\end{align}
and let $\vec{a}^x_W,\vec{a}^y_W\in\mathcal{M}_W$. Then for any $0\leq\theta\leq 1$, we have $\vec{a}^z_W=\theta\vec{a}^x_W+(1-\theta)\vec{a}^y_W\in\mathcal{M}_W$, which means the set $\mathcal{M}_W$ is convex. This can be seen by
\begin{align*}
g(\vec{a}^z_W)&=\theta g(\vec{a}^x_W)+(1-\theta)g(\vec{a}^y_W)=1,\\ \|\vec{a}^z_W\|_{\infty}&\leq\theta\|\vec{a}^x_W\|_{\infty}+(1-\theta)\|\vec{a}^y_W\|_{\infty}\leq d_W.
\end{align*}
Since $\mathcal{M}_W$ is closed and bounded in the finite dimensional normed space ${\mathbb R}^{W}$, it is compact. Moreover, for any $\vec{a}^0_W\in\mathcal{M}_W$, $\hat{\mathcal{F}}(\vec{a}_W)\rightarrow\hat{\mathcal{F}}(\vec{a}^0_W)$ as $\vec{a}_W\rightarrow\vec{a}^0_W$. Thus, $\hat{\mathcal{F}}(\vec{a}_W)$ is continuous on $\mathcal{M}_W$.
According to \textit{Proposition \ref{ProAwBounded}}, we have $\hat{\mathcal{F}}(\mathcal{M}_W)\subseteq\mathcal{M}_W$, where $\hat{\mathcal{F}}(\mathcal{M}_W)$ is the images of $\mathcal{M}_W$. Now $\hat{\mathcal{F}}$ is a continuous map of the nonempty, convex, compact set $\mathcal{M}_W$ into itself. By \textit{Lemma \ref{LemFixedPointBrouwer}}, there exists a fixed point for $\hat{\mathcal{F}}$ and consequently the nonlinear equations \eqref{EquOptimalWeightFstNesCond} have a solution.\qed


\subsection{Symmetry of the optimal weight}
Intuitively, since the process is stationary, the first and last samples in a time window always have the same contributions to the covariance matrix $\tilde\Sigma_W$, as can be seen in \eqref{WeightedSampleDistribution}. Therefore, they should have the same weight when $N$ is sufficiently large. This is also true for the second and the penultimate samples, and so on. In this subsection, we reveal that the optimal weight possesses a symmetrical structure, see \textit{Theorem \ref{ThmSymmetryAW}}.
\begin{proposition}\label{ProRijSwInfi}
When $N$ is sufficiently large, we have
\begin{align}
\label{EquRijInfi}
{\mathbb E}(\hat{R}_{lj})=R_{l-j},\quad&\lim_{N\to\infty}\hat{R}_{lj}=R_{l-j}, a.s. \\
\label{EquSwInfi}
{\mathbb E}(\tilde{S}_W)=\tilde{\Sigma}_W,\quad&\lim_{N\to\infty}\tilde{S}_W=\tilde{\Sigma}_W, a.s.
\end{align}
\end{proposition}
\textbf{Proof.} According to \eqref{SwHatR}, we can derive \eqref{EquSwInfi} directly if \eqref{EquRijInfi} holds. As for \eqref{EquRijInfi}, note that $X^{j_1}_{i_1}$ and $X^{j_2}_{i_2}$ are independent for $i_1\!\neq\!i_2$. Thus, ${\mathbb E}(X^l_i(\bar{X}^j)^T)={\mathbb E}(\bar{X}^l (X^j_i)^T)=\frac{1}{N}R_{l-j}+\mu\mu^T$ and ${\mathbb E}(\bar{X}^l(\bar{X}^j)^T)=\frac{1}{N}R_{l-j}+\mu\mu^T$. Then
\begin{align*}
{\mathbb E}(\hat{R}_{lj})=\frac{1}{N-1}\sum\limits_{i=1}^{N}{\mathbb E}&\left[X^l_i(X^j_i)^T-X^l_i(\bar{X}^j)^T\right.\\
&-\left.\bar{X}^l(X^j_i)^T+\bar{X}^l(\bar{X}^j)^T\right]=R_{l-j}.
\end{align*}
Moreover, it is well known that the stationary Gaussian process is ergodic. Thus,
\begin{align*}
&\lim_{N\to\infty}\bar{X}^j=\lim_{N\to\infty}\frac{1}{N}\sum\limits_{i=1}^{N}X^j_i={\mathbb E}(X^j_i)=\mu,\;a.s.,\\ &\lim_{N\to\infty}\frac{1}{N\!-\!1}\sum\limits_{i=1}^{N}X^l_i(X^j_i)^T\!=\!{\mathbb E}\left(X^l_i(X^j_i)^T\right)\!=\!R_{l-j}+\mu\mu^T,a.s.
\end{align*}
Substituting them into \eqref{EquOptimalWeightFstNesCondSpt}, we derive \eqref{EquRijInfi}. \qed

\begin{theorem}\label{ThmSymmetryAW}
Suppose \textit{Assumption \ref{AssmGammaW}} holds and $N$ is sufficient large, then the optimal weight $\vec{a}^*_W$ maximizing $\beta(\vec{a}_W)$ of Problem \ref{PlmOptimalWeight} satisfies
\begin{align}\label{EquSymmetryAW}
a^*_j=a^*_{W-j+1},\quad 1\leq j\leq W.
\end{align}
\end{theorem}
\textbf{Proof.} According to \textit{Proposition \ref{ProRijSwInfi}}, when $N$ is sufficiently large, we almost surely have
\begin{align*}
\mathcal{T}_{l,j}&=\left\{\begin{array}{ll}
\xi^T_q\tilde{\Sigma}_W^{-1}\left(R_{l-j}-R_{l+1-j}\right)\tilde{\Sigma}_W^{-1}\xi_q,&{}\quad l<W,\\
1,&{}\quad l=W,
\end{array}\right.
\end{align*}
where $\mathcal{T}$ is short for $\mathcal{T}(\vec{a}_W)$, such that
\begin{align}\label{EquOptimalWeightInfiFstNesCond}
\mathcal{T}&(\vec{a}^*_W)\vec{a}^*_W=b.
\end{align}
Following $R_{-l}=R^T_{l}$, it can be seen that
\begin{align*}
\begin{array}{ll}
\mathcal{T}_{l,j}=\mathcal{T}_{l+1,j+1},&{}\quad l<W-1, j<W,\\
\mathcal{T}_{l,j}=-\mathcal{T}_{j-1,l},&{}\quad l\leq W-1, j\leq W.
\end{array}
\end{align*}
Denote $A_{\backslash l\backslash\emptyset}$ and $A_{\backslash\emptyset\backslash j}$ as the matrices obtained from $A$ by deleting the $l$th row and $j$th column, respectively. Then, $\mathcal{T}_{\backslash W\backslash\emptyset}\in{\mathbb R}^{W-1\times W}$ has the following form
\begin{align*}
\mathcal{T}_{\backslash W\backslash\emptyset}&=\left[ {\begin{array}{cccccc}
-t_1 & t_1 & t_2 & \cdots & t_{W-2} & t_{W-1}\\
-t_2 & -t_1 & t_1 & t_2 & \ddots & t_{W-2}\\
-t_3 & -t_2 & -t_1 & t_1 & \ddots & \vdots\\
\vdots & \ddots & \ddots & \ddots & \ddots & t_2\\
-t_{W-1} & \cdots & -t_3 & -t_2 & -t_1 & t_1\\
\end{array}} \right],
\end{align*}
where $t_l=\xi^T_q\tilde{\Sigma}_W^{-1}\left(R_{-l}-R_{1-l}\right)\tilde{\Sigma}_W^{-1}\xi_q$, and is an abbreviation for $t_l(\vec{a}_W)$.
Define $\textrm{cen}(A)\in{\mathbb R}^{m\times n}$ as the centrosymmetry of a matrix $A\in{\mathbb R}^{m\times n}$, namely, $[\textrm{cen}(A)]_{l,j}\!=\![A]_{m-l+1,n-j+1}$.
It can be easily verified that the operator $\textrm{cen}()$ has the following properties:
\begin{align}\label{CenProperty1}
\textrm{cen}\left(\textrm{cen}(A)\right)=A,\quad\textrm{cen}(-A)=-\textrm{cen}(A).
\end{align}
Besides, if $A$ is a square matrix, then we have $|\textrm{cen}(A)|=|A|$.
Moreover, if $A$ is centrosymmetric, that is to say, $\textrm{cen}(A)=A$, then we have
\begin{align}\label{CenProperty2}
\textrm{cen}(A_{\backslash\emptyset\backslash j})=A_{\backslash\emptyset\backslash n-j+1}.
\end{align}
Note that $[\mathcal{T}_{\backslash W\backslash\emptyset}]_{l,j}\!=\!-[\mathcal{T}_{\backslash W\backslash\emptyset}]_{W-l,W-j+1}$, i.e. $\textrm{cen}(\mathcal{T}_{\backslash W\backslash\emptyset})\!=\!-\mathcal{T}_{\backslash W\backslash\emptyset}$. According to \eqref{CenProperty1} and \eqref{CenProperty2}, we have
\begin{align}
\textrm{cen}(\mathcal{T}_{\backslash W\backslash j})=-\mathcal{T}_{\backslash W\backslash W-j+1}.
\end{align}
Thus,
\begin{align*}
|\mathcal{T}_{\backslash W\backslash j}|=|-\textrm{cen}(\mathcal{T}_{\backslash W\backslash W-j+1})|=(-1)^{W-1}|\mathcal{T}_{\backslash W\backslash W-j+1}|.
\end{align*}
Since $\mathcal{T}(\vec{a}_W)$ is nonsingular for any weight vector $\vec{a}_W\neq 0_W$, similar to the Cramer's rule, we have $\vec{a}^*_W=\mathcal{T}^{-1}(\vec{a}^*_W)b=|\mathcal{T}(\vec{a}^*_W)|^{-1}\textrm{adj}\left(\mathcal{T}(\vec{a}^*_W)\right)b$. Then
\begin{align*}
a^*_j&=(-1)^{W+j}|\mathcal{T}(\vec{a}^*_W)|^{-1}|\mathcal{T}_{\backslash W\backslash j}(\vec{a}^*_W)|\nonumber\\
&=(-1)^{2W-j+1}|\mathcal{T}(\vec{a}^*_W)|^{-1}|\mathcal{T}_{\backslash W\backslash W-j+1}(\vec{a}^*_W)|=a^*_{W-j+1},
\end{align*}
which completes the proof. \qed

\subsection{Further results in several special cases}
Note that \textit{Theorem \ref{ThmOptimalWeight}} only gives the necessary conditions.
Nevertheless, in some special cases, we can further find necessary and sufficient conditions, and determine the optimal weight exactly.

\begin{proposition}\label{ProOptimalWeightsufficient}
If $\vec{a}^*_W$ meets \eqref{EquOptimalWeightFstNesCond} and makes the strict inequality in \eqref{EquOptimalWeightSecNesCond} hold, then it is a maximum point of $\beta(\vec{a}_W)$ in Problem \ref{PlmOptimalWeight}.
\end{proposition}
\textbf{Proof.} According to \cite{Chiang2005Fundamental}, when $\vec{a}^*_W$ meets the first-order necessary conditions  \eqref{EquOptimalWeightFstNesCond}, the second-order sufficient conditions for $\vec{a}^*_W$ to be a maximum point (rather than a minimum or saddle point) are: the leading principal minors of $\overline{{\cal H}}(\vec{a}^*_W)$ of order $k+\!1$ ($k\!=\!2,3,\!\cdots\!,\!W$) have sign $(-1)^{k}$. By following \textit{Theorem \ref{ThmOptimalWeight}}, we obtain this proposition. \qed

\begin{proposition}\label{ProSimplifybeta}
For the optimal weight $\vec{a}^*_W$ maximizing $\beta(\vec{a}_W)$ of Problem \ref{PlmOptimalWeight}, we have
\begin{align}\label{EquSimplifybeta}
2\beta(\vec{a}^*_W)=\hat{\gamma}^{W}_{l,:}(\vec{a}^*_W)\vec{a}^*_W,\quad 1\leq l\leq W.
\end{align}
\end{proposition}
\textbf{Proof.} It follows from \eqref{SwHatR} that
\begin{align*}
\left\|{\tilde{S}_W^{-1/2}\xi_q}\right\|^2=\xi^T_q\tilde{S}^{-1}_W&\left(\vec{a}_W\otimes I_p\right)^T\hat{\Gamma}_W\left(\vec{a}_W\otimes I_p\right)\tilde{S}^{-1}_W\xi_q\nonumber\\
=\left(\vec{a}_W\otimes \tilde{S}^{-1}_W\xi_q\right)^T&\hat{\Gamma}_W\left(\vec{a}_W\otimes \tilde{S}^{-1}_W\xi_q\right)\nonumber\\
=\vec{a}^T_W\left(I_W\otimes\tilde{S}^{-1}_W\xi_q\right)^T&\hat{\Gamma}^W\left(I_W\otimes\tilde{S}^{-1}_W\xi_q\right)\vec{a}_W
=\vec{a}^T_W\hat{\gamma}^{W}\vec{a}_W.
\end{align*}
Note that we have $\hat{\gamma}^{W}_{l,:}(\vec{a}^*_W)\vec{a}^*_W=\hat{\gamma}^{W}_{l',:}(\vec{a}^*_W)\vec{a}^*_W$ from \eqref{EquOptimalWeightFstNesCond}. Thus,
\begin{align*}
&2\beta(\vec{a}^*_W)=\left\|{\tilde{S}_W^{-1/2}(\vec{a}^*_W)\xi_q}\right\|^2=(\vec{a}^*_W)^T\hat{\gamma}^{W}(\vec{a}^*_W)\nonumber\\
&=(\vec{a}^*_W)^T[\hat{\gamma}^{W}_{:,l}(\vec{a}^*_W),\cdots,\hat{\gamma}^{W}_{:,l}(\vec{a}^*_W)]\vec{a}^*_W
=(\vec{a}^*_W)^T\hat{\gamma}^{W}_{:,l}(\vec{a}^*_W).
\end{align*}
The last equality is because $1^T_W\vec{a}^*_W=1$. Then by following $(\hat{\gamma}^{W}_{:,l})^T=\hat{\gamma}^{W}_{l,:}$, we obtain \eqref{EquSimplifybeta}. \qed

\begin{theorem}\label{ThmOptimalWeightIndp}
When process data are independent, $R_0$ is nonsingular and $N$ is sufficiently large, the optimal weight $\vec{a}^*_W$ maximizing $\beta(\vec{a}_W)$ of Problem \ref{PlmOptimalWeight} is uniquely determined as
\begin{align}\label{EquOptimalWeightIndp}
a^*_1=a^*_2=\cdots=a^*_W=1/W.
\end{align}
\end{theorem}
\textbf{Proof.} When process data are independent, we have $R_l=0$, $\forall l\neq 0$. By following the proof of \textit{Theorem \ref{ThmSymmetryAW}}, when $N$ is sufficiently large, we almost surely have
\begin{align}
\mathcal{T}(\vec{a}_W)=\left[ \begin{array}{c}
\begin{array}{ccccc}
t_0 & -t_0 & 0 & \cdots & 0\\
0 & t_0 & -t_0 & \ddots & \vdots\\
\vdots & \ddots & \ddots & \ddots & 0\\
0 & \cdots & 0 & t_0 & -t_0\\
1 & \cdots & 1 & \cdots & 1\end{array} \\
\end{array}\right],
\end{align}
where $t_0=\xi^T_q\tilde{\Sigma}_W^{-1}R_{0}\tilde{\Sigma}_W^{-1}\xi_q>0$. Note that
\begin{align*}
|\mathcal{T}(\vec{a}_W)|=W(t_0)^{W-1},\ |\mathcal{T}_{\backslash W\backslash j}(\vec{a}_W)|=(-1)^{W-j}(t_0)^{W-1}.
\end{align*}
Thus, $a^*_j=(-1)^{W+j}|\mathcal{T}(\vec{a}^*_W)|^{-1}|\mathcal{T}_{\backslash W\backslash j}(\vec{a}^*_W)|=1/W$. Then, $\tilde\Sigma_W(\vec{a}^*_W)=\frac{1}{W}R_0$. Substituting them into \eqref{EquOptimalWeightSecNesCondSpt}, we have $\hat{H}_{l,l'}(\vec{a}^*_W)\rightarrow H_{l,l'}(\vec{a}^*_W)=4W\vartheta-W^2\vartheta\delta_{ll'}$ almost surely when $N$ is sufficiently large, where $\vartheta=\xi^T_qR^{-1}_0\xi_q$. Thus,
\begin{align*}
\left| \overline{{\cal H}}_{k}(\vec{a}^*_W) \right|=
\left| {\begin{array}{cc}
0 & 1^T_k\\
1_k & -(W^2\vartheta)I_k
\end{array}} \right|
=k(-1)^{k}(W^2\vartheta)^{k-1}.
\end{align*}
Following \textit{Proposition \ref{ProOptimalWeightsufficient}}, \eqref{EquOptimalWeightIndp} is the optimal weight. \qed

\begin{theorem}\label{ThmEqualWeightNesCond}
Suppose \textit{Assumption \ref{AssmGammaW}} holds and $N$ is sufficient large, if the optimal weight $\vec{a}^*_W$ maximizing $\beta(\vec{a}_W)$ of Problem \ref{PlmOptimalWeight} is \eqref{EquOptimalWeightIndp}, then
\begin{align}\label{EquEqualWeightNesCond}
\xi^T_q\tilde{\Sigma}_W^{-1}(\!\frac{1_W}{W}\!)\left(\!{R}_{j}\!-\!{R}_{W-j}\!\right)\tilde{\Sigma}_W^{-1}(\!\frac{1_W}{W}\!)\xi_q\!=\!0,1\!\leq\!j\!\leq\!W\!-\!1.
\end{align}
\end{theorem}
\textbf{Proof.} By substituting \eqref{EquOptimalWeightIndp}, i.e., $\vec{a}^*_W=\frac{1_W}{W}$, into \eqref{EquOptimalWeightInfiFstNesCond}, we have
\begin{align*}
\sum\limits_{l=j+1}^{W-j}t_l(\frac{1_W}{W})=0,\ j=1,2,\cdots,[\frac{W\!-\!1}{2}]^-,
\end{align*}
which is equivalent to \eqref{EquEqualWeightNesCond}. Here, $[x]^-$ represents the maximum integer no more than $x$. \qed

\begin{remark}
\textit{Theorem \ref{ThmOptimalWeightIndp}} explains why the MA scheme, i.e., the equally weighted scheme, is always adopted in FD tasks where samples are assumed to be independent, such as in \cite{Chen2001Principle,Ji2016Incipient,Ji2017Incipient}. Note that when process data are independent, \eqref{EquEqualWeightNesCond} holds since $R_l=0$, $\forall l\neq 0$. \textit{Theorem \ref{ThmEqualWeightNesCond}} further gives a necessary condition for the MA scheme to be optimal. The essence that the equally weighted scheme does not effectively utilize the correlation information of samples is revealed here. When $p=1$, \eqref{EquEqualWeightNesCond} becomes a necessary and sufficient condition and is equivalent to ${R}_{j}={R}_{W-j}$, $1\!\leq\!j\!\leq\!W\!-\!1$. This means that in the unidimensional case, the MA scheme is optimal only for special stationary processes that have periodicity.
\end{remark}

\begin{theorem}\label{ThmOptimalWeightUnidm}
When $p=1$, suppose \textit{Assumption \ref{AssmGammaW}} holds, then the optimal weight $\vec{a}^*_W$ maximizing $\beta(\vec{a}_W)$ of Problem \ref{PlmOptimalWeight} is uniquely determined as
\begin{align}\label{EquOptimalWeightUnidm}
\vec{a}^*_W=\hat{A}^{-1}b,
\end{align}
where
\begin{align}
\hat{A}_{l,j}=\left\{\begin{array}{ll}
\hat{R}_{lj}-\hat{R}_{(l+1)j},&{}\quad l<W,\\
1,&{}\quad l=W.
\end{array}\right.
\end{align}
\end{theorem}
\textbf{Proof.} When $p=1$, we have $\xi_q=1$, and $\tilde{S}_W(\vec{a}_W)$ is a scalar. Thus, \eqref{EquOptimalWeightFstNesCond} degenerates into linear equations with unique solutions \eqref{EquOptimalWeightUnidm}. It follows from \textit{Proposition \ref{ProSimplifybeta}} that
\begin{align}
2\beta(\vec{a}^*_W)=\tilde{S}^{-1}_W(\vec{a}^*_W)=\hat{\gamma}^{W}_{l,:}(\vec{a}^*_W)\vec{a}^*_W,\quad 1\leq l\leq W.
\end{align}
Multiplying both sides by $\tilde{S}^2_W(\vec{a}^*_W)$ and following \eqref{gammaHatR}, we obtain $\tilde{S}_W(\vec{a}^*_W)=\hat{\Gamma}^W_{l,:}\vec{a}^*_W$. Substituting it into \eqref{EquOptimalWeightSecNesCondSpt}, we have
\begin{align*}
\hat{H}_{l,l'}(\vec{a}^*_W)&=4\tilde{S}^{-1}_W(\vec{a}^*_W)-\tilde{S}^{-2}_W(\vec{a}^*_W)\hat{R}_{ll'},
\end{align*}
and thus,
\begin{align*}
\left| \overline{{\cal H}}_{k}(\vec{a}^*_W) \right|&=
\left| {\begin{array}{cc}
0 & 1^T_k\\
1_k & \tilde{S}^{-2}_W(\vec{a}^*_W)\hat{\Gamma}^k
\end{array}} \right|\nonumber\\
&=(-1)^{k}\left(\tilde{S}^{2}_W(\vec{a}^*_W)1^T_k(\hat{\Gamma}^k)^{-1}1_k\right)\left|\tilde{S}^{-2}_W(\vec{a}^*_W)\hat{\Gamma}^k\right|.
\end{align*}
Note that $\tilde{S}_W, \hat{\Gamma}^k$ are positive definite, then
\begin{align*}
(-1)^{k}\left| \overline{{\cal H}}_{k}(\vec{a}^*_W) \right|>0,\quad k=2,3,\cdots,W.
\end{align*}
Following \textit{Proposition \ref{ProOptimalWeightsufficient}}, \eqref{EquOptimalWeightUnidm} is the optimal weight. \qed

\begin{theorem}\label{ThmOptimalWeightTwo}
When $W=2$, suppose \textit{Assumption \ref{AssmGammaW}} holds and $N$ is sufficiently large, then the optimal weight $\vec{a}^*_W$ maximizing $\beta(\vec{a}_W)$ of Problem \ref{PlmOptimalWeight} is uniquely determined as
\begin{align}\label{EquOptimalWeightTwo}
a^*_1=a^*_2=1/2.
\end{align}
\end{theorem}
\textbf{Proof.} When $W=2$, \eqref{EquOptimalWeightTwo} can be derived directly from \textit{Theorem \ref{ThmSymmetryAW}}. When $N$ is sufficiently large, we have
\begin{align*}
\hat{h}_{l}(\vec{a}_W)\rightarrow h_{l}(\vec{a}_W)=\tilde{\Sigma}_W^{-1/2}\left(\sum\limits_{j=1}^{W}a_j(R_{l-j}+R^T_{l-j})\right)\tilde{\Sigma}_W^{-1}\xi_q.
\end{align*}
Note that when $W=2$, we have $h_1(\vec{a}^*_2)=h_2(\vec{a}^*_2)$. Thus,
\begin{align*}
\left| \overline{{\cal H}}_{k}(\vec{a}^*_2) \right|=
\left| {\begin{array}{cc}
0 & 1^T_k\\
1_k & -\hat{\gamma}^k
\end{array}} \right|
=(-1)^{k}\left(1^T_k(\hat{\gamma}^k)^{-1}1_k\right)|\hat{\gamma}^k|.
\end{align*}
Following \textit{Proposition \ref{ProOptimalWeightsufficient}}, \eqref{EquOptimalWeightTwo} is the optimal weight. \qed

\begin{remark}
Note that the derived weights \eqref{EquOptimalWeightIndp}, \eqref{EquOptimalWeightUnidm} and \eqref{EquOptimalWeightTwo} are optimal regardless of the direction of IFs in these three cases, respectively.
\end{remark}

\subsection{Selection of the window length}
One drawback of introducing a time window is that it causes detection delays. Generally speaking, an overlarge window length may incur serious detection delays. As a result, we suggest choosing the smallest window length that guarantees the detection of IFs.

\begin{theorem}\label{ThmIFD1}
For the OWMA-TCC with $W\!\leq\!W^\#$ and a given significance level $\alpha$, the $q$th IF is guaranteed detectable {(G-detectable)} if and only if
\begin{align}
\beta(\vec{a}^{*}_{W^\#})f^2_q>2\delta^2.
\end{align}
Then the window length $W$ can be chosen such that $W^\#\geq W\geq W^*$, where
\begin{align}\label{EquThmIFD1OptW}
W^*=\arg\min_{W}\ \beta(\vec{a}^{*}_W)f^2_q>2\delta^2.
\end{align}
\end{theorem}
\textbf{Proof.} Note that $\beta([\vec{a}^*_{W-1};0])\leq\beta(\vec{a}^*_W)$. Thus, we can conclude that when $W\!\leq\!W^\#$, the maximum of $\beta(\vec{a}_W)$ achieves with $W=W^\#$ and $\vec{a}_W=\vec{a}^*_W$.
Then, this theorem holds according to \textit{Theorem \ref{ThmIFD0}}. \qed

In \eqref{EquThmIFD1OptW}, $W^*$ can be solved by exhaustive search from $W=1$ to $W^\#$. Since larger window length always incurs larger detection delays, we can select $W^*$ as the optimal window length and $\vec{a}^{*}_{W^*}$ as the optimal weight vector. In practice, IFs' parameters may not be know exactly, but in most cases lower bounds of fault parameters are available through expert knowledge or analyzing historical data and operating conditions. Denote $\tilde{f}_q,\tilde{\tau}^r_{q-1},\tilde{\tau}^o_q,\tilde{\tau}^r_{q}$ as the lower bounds of $f_q,\tau^r_{q-1},\tau^o_q,\tau^r_{q}$, respectively. Then we have the following corollaries.

\begin{corollary}\label{ColIFDlow0}
For the WMA-TCC($W$) and a given significance level $\alpha$, when $W\!\leq\!\tilde{W}^\#\!\triangleq\!\min\{\tilde{\tau}^r_{q-1},\tilde{\tau}^o_{q},\tilde{\tau}^r_{q}\}$, the $q$th IF is guaranteed detectable {(G-detectable)} if IF$(\xi_q,\tilde{f}_q,\tilde{\tau}^r_{q-1},\tilde{\tau}^o_q,\tilde{\tau}^r_{q})$ is guaranteed detectable.
\end{corollary}
\textbf{Proof.} Directly derived from \textit{Theorem \ref{ThmIFD0}}. \qed

\begin{corollary}\label{ColIFDlow1}
For the OWMA-TCC with $W\!\leq\!\tilde{W}^\#$ and a given significance level $\alpha$, the $q$th IF is guaranteed detectable {(G-detectable)} if
\begin{align}
\beta(\vec{a}^{*}_{\tilde{W}^\#})\tilde{f}^2_q>2\delta^2.
\end{align}
Then the window length $W$ can be chosen such that $\tilde{W}^\#\geq W\geq \tilde{W}^*$, where
\begin{align}
\tilde{W}^*=\arg\min_{W}\ \beta(\vec{a}^{*}_W)\tilde{f}^2_q>2\delta^2.
\end{align}
\end{corollary}
\textbf{Proof.} Directly derived from \textit{Theorem \ref{ThmIFD1}} and \textit{Corollary \ref{ColIFDlow0}}. \qed

\begin{remark}
A PF can be viewed as an IF with an infinite active duration. Thus, the developed methods as well as the above analyses, including all the theorems, propositions and corollaries, are applicable to PF by setting $\tau^r_{q-1},\tau^o_q,\tau^r_{q}\to\infty$. Moreover, the developed OWMA method can be combined with dimensionality reduction techniques such as PCA and PLS to monitor specific subspaces, by replacing the measurement vector $X$ with its score vector in corresponding subspaces.
\end{remark}

\section{Generalization to weakly stationary processes without the Gaussianity assumption}\label{NonGaussianSec}

This section extends the above results to weakly stationary processes without the Gaussianity assumption. It can be seen from the previous sections that to implement the developed OWMA method, the autocovariance function $R_l$ of the stationary process is needed, instead of the exact distribution of the stationary process. Thus, the developed OWMA method can be used in any stationary process with or without the Gaussianity assumption. Then, the remained question is to prove the optimality of the developed OWMA method in general stationary processes without Gaussianity assumption. Note that the stationary process here means the weakly stationary process, unless we specifically indicate otherwise.

It is worth pointing out that the $T^2$ statistic is widely adopted under both Gaussian and non-Gaussian conditions. It is well-known in the statistics \cite{Anderson2003An} that under the condition of Gaussian distribution, the $T^2$ test is the uniformly most powerful unbiased test (UMPUT) of the hypothesis $H_0: \mu_f=\mu$ versus $H_1: \mu_f\neq\mu$. Therefore, the $T^2$ statistic has been widely accepted for hypothesis testing problem under the Gaussian condition. Later, due to its simple form and its optimal properties under the Gaussian condition, the $T^2$ statistic has been also widely adopted to detect anomalies under non-Gaussian conditions. In these cases, the $T^2$ statistic is understood as a measure of the process variation, and thus can be used in the process monitoring task under non-Gaussian conditions. Here, a little difference between the use of OWMA-TCC in stationary processes with and without the Gaussianity assumption is the calculation of the control limit. Without the Gaussianity assumption, the WMA-$T^2$ statistic does not follow an $F$ distribution, and thus the control limit can not be calculated by \eqref{UclConsecutive}. Corresponding solutions are to use the empirical method \cite{Shang2017Recursive} or the kernel density estimation (KDE) method \cite{Pilario2018Canonical} to calculate the control limit of $\tilde{T}^2_k(W)$, denoted as $\delta^2_W$, under non-Gaussian conditions.

Now we are in the position to show that, there is a sense in which the weight vector $\vec{a}^*_W$ defined by \textit{Problem \ref{PlmOptimalWeight}} and subsequently given in \textit{Theorem \ref{ThmOptimalWeight}} is also optimal for general weakly stationary processes. We consider that the stationary process is ergodic and $N$ is sufficiently large. In this case, we have
\begin{align}\label{T2ConseInfi}
\lim_{N\to\infty}\tilde{T}^2_k(W)=(\tilde{X}^f_k-\mu)^T\tilde{\Sigma}^{-1}_W(\tilde{X}^f_k-\mu),
\end{align}
and
\begin{align}\label{T2ConseInfiMean}
{\mathbb E}&\left\{\lim_{N\to\infty}\tilde{T}^2_k(W)\right\}={\mathbb E}\left\{{\mathrm {tr}}[(\tilde{X}^f_k-\mu)(\tilde{X}^f_k-\mu)^T\tilde{\Sigma}^{-1}_W]\right\}\nonumber\\
&={\mathrm {tr}}\left\{{\mathbb E}[(\tilde{X}^f_k-\mu)(\tilde{X}^f_k-\mu)^T]\tilde{\Sigma}^{-1}_W\right\},
\end{align}
where $\tilde{X}^f_k$ is modeled by \eqref{FaultModel}, \eqref{FaultModelWin} and \eqref{IFModel}.
Then, the mean of the WMA-$T^2$ statistic under the hypothesis of no fault, i.e., $H_0$, is
\begin{align}\label{T2ConseInfiMeanH0}
{\mathbb E}&\left\{\lim_{N\to\infty}\tilde{T}^2_k(W){\Big |}H_0\right\}\\
&={\mathrm {tr}}\left\{{\mathbb E}[(\tilde{X}^*_k-\mu)(\tilde{X}^*_k-\mu)^T]\tilde{\Sigma}^{-1}_W\right\}=p,\nonumber
\end{align}
where the last equality is because the mean and covariance matrix of $\tilde{X}^*_k$ are $\mu$ and $\tilde\Sigma_W$, respectively. In addition, the mean of the WMA-$T^2$ statistic under the hypothesis of IFs, i.e., $H_1$, is
\begin{align}\label{T2ConseInfiMeanH1}
{\mathbb E}&\left\{\lim_{N\to\infty}\tilde{T}^2_k(W){\Big |}H_1\right\}\\
&={\mathrm {tr}}\left\{{\mathbb E}[(\tilde{X}^*_k-\mu+\xi_q{f_q})(\tilde{X}^*_k-\mu+\xi_q{f_q})^T]\tilde{\Sigma}^{-1}_W\right\}\nonumber\\
&={\mathrm {tr}}\left\{[\tilde\Sigma_W+\xi_q\xi^T_q{f^2_q}]\tilde{\Sigma}^{-1}_W\right\}=\left\|{\tilde{\Sigma}_W^{-1/2}\xi_q{f_q}}\right\|^2+p.\nonumber
\end{align}
By following \eqref{T2ConseInfiMeanH0} and \eqref{T2ConseInfiMeanH1}, it can be seen that
\begin{align*}
&{\mathbb E}\left\{\lim_{N\to\infty}\tilde{T}^2_k(W){\Big |}H_1\right\}-{\mathbb E}\left\{\lim_{N\to\infty}\tilde{T}^2_k(W){\Big |}H_0\right\}\\
&=\left\|{\tilde{\Sigma}_W^{-1/2}\xi_q{f_q}}\right\|^2=\lim_{N\to\infty}2\beta(\vec{a}_{W})f^2_q,
\end{align*}
where $\beta(\vec{a}_{W})$ is the objective function of \textit{Problem \ref{PlmOptimalWeight}}.

For a hypothesis testing problem, one always prefers the distribution of the designed statistic to be as different as possible between two hypotheses. To this end, a widely used tool is the KL divergence. The KL divergence is designed to measure the difference between two probability density functions (PDFs) \cite{Zeng2014Detecting}. However, calculation of the KL divergence needs exact PDFs of the distributions under two hypotheses. An alternative way is to measure the difference between two PDFs through the distance of their means, such as in \cite{Chen2001Principle}. Note that $f_q$ is not a function of $\vec{a}_{W}$. Therefore, $\beta(\vec{a}_{W})$ can measure the distribution difference of the WMA-$T^2$ statistic between two hypotheses in stationary processes. That is to say, the developed OWMA is optimal for weakly stationary processes in the sense of the statistic's distribution difference between two hypotheses.

\begin{remark}
To sum up, the use of $T^2$ statistic is reasonable under both Gaussian and non-Gaussian conditions. Moreover, the developed OWMA-TCC is optimal for weakly stationary processes in the sense of the $T^2$ statistic's distribution difference between two hypotheses. To implement the developed OWMA method under non-Gaussian conditions, the only difference is to use the empirical method or the KDE method to calculate the control limit $\delta^2_W$. Note that the detectability analyses conducted in Section \ref{DetectabilitySec} do not employ the Gaussianity assumption. Hence, by replacing the control limit $\delta$ therein with $\delta_W$, they become valid for weakly stationary processes. Properties of the OWMA method given in Section \ref{OptimalWeightKnownXiSec} are as well valid for weakly stationary processes.
\end{remark}

\section{Simulation studies}\label{SimulationSec}
In this section, two simulation examples are used to demonstrate the efficiency of the OWMA-TCC under both Gaussian and non-Gaussian conditions, by comparing with existing static and dynamic MSPM methods.
\subsection{A numerical example}\label{NumSimuSub}
A multivariate AR($1$) process model used in the original DPCA literature \cite{Ku1995Disturbance} is employed here to illustrate the effectiveness and efficiency of the developed method, in comparison with several well-known methods. The process model under normal operating conditions is
\begin{align}\label{NumericalModelY}
\textbf{z}_k&=\begin{bmatrix}0.118&-0.191 \\0.847&0.264 \\ \end{bmatrix}\textbf{z}_{k-1}+ \begin{bmatrix}1&2 \\3&-4 \\ \end{bmatrix}\textbf{u}_{k-1},\nonumber\\
\textbf{y}_k&=\textbf{z}_k+\textbf{v}_k,
\end{align}
where $\textbf{u}$ is the correlated input:
\begin{align*}
\textbf{u}_k&=\begin{bmatrix}0.811&-0.226 \\0.477&0.415 \\ \end{bmatrix}\textbf{u}_{k-1}+ \begin{bmatrix}0.193&0.689 \\-0.320&-0.749 \\ \end{bmatrix}\textbf{w}_{k-1}.
\end{align*}
According to \cite{Ku1995Disturbance}, the noises $\textbf{w}$ and $\textbf{v}$ are zero means, and follow Gaussian distributions with variance $1$ and $0.1$, respectively. Both $\textbf{u}$ and $\textbf{y}$ are measured so that we can form the process data as $X_k=[\textbf{y}_k;\textbf{u}_k]$.

\begin{figure}
\begin{center}
\includegraphics[width=7.7cm]{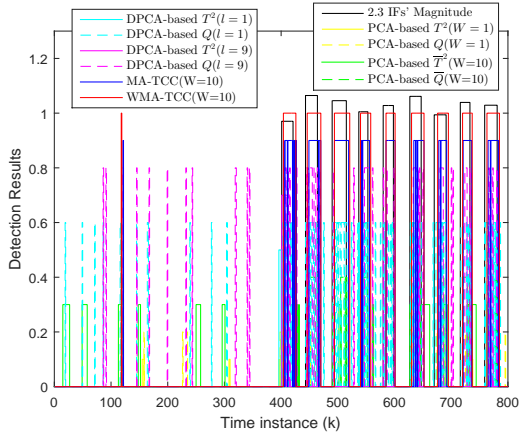}    
\caption{IFD results using different methods in the numerical simulation (Gaussian noise).}
\label{DetectionResultsAllNum}
\end{center}
\end{figure}

Both 5000 sets of 10 consecutive observations (training samples) and 800 consecutive observations (test samples) are generated according to (\ref{NumericalModelY}), and intermittent process faults are subsequently introduced in the test dataset since sample 401. The first 400 test samples are used to calculate false alarm rates (FARs) of different methods. The introduced IFs have an additive form as modeled by (\ref{IFModel}) with the fault direction $\xi_q\!=\![0.0319,-0.2740,0.9611,-0.0098]^T$, the lower bound of each fault magnitude $\tilde{f}_q\!=\!0.42$, the lower bound of each fault active and inactive duration $\tilde{\tau}^o_q\!=\!15,\tilde{\tau}^r_{q}\!=\!20$. The actual fault magnitude, fault active and inactive duration are all generated randomly according to their lower bounds and are shown in Fig.~\ref{DetectionResultsAllNum} with a black line (the $Y$-axis shows the fault magnitude multiplied by 2.3, and the $X$-axis shows the fault active and inactive duration).

\begin{figure}
\begin{center}
\includegraphics[width=7.7cm]{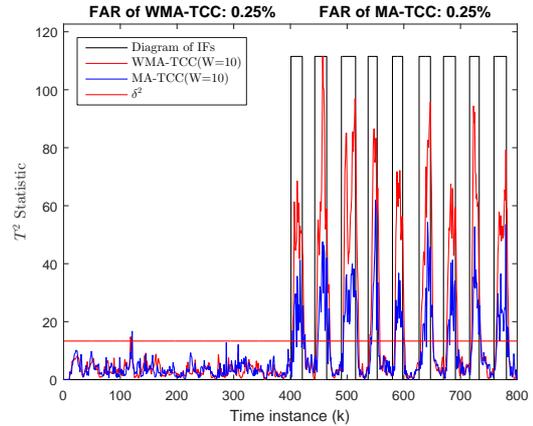}    
\caption{IFD using the OWMA-TCC and MA-TCC with window length $W\!=\!10$ in the numerical simulation (Gaussian noise).}
\label{WMAaMANum}
\end{center}
\end{figure}

Training samples are used to determine the optimal weight vector and the significance level $\alpha$ is set as $0.01$. Then, we can conclude that the introduced IFs are guaranteed detectable by the OWMA-TCC with window length $W\!\in\![10,15]$, according to \textit{Theorems \ref{ThmIFD0}} and \ref{ThmIFD1} and \textit{Corollary \ref{ColIFDlow1}}. The OWMA-TCC with window length $W\!=\!10$ is given in Fig.~\ref{WMAaMANum} with a red line. To demonstrate the importance of employing an optimal weight vector, the WMA-TCC with the equally weighted scheme, denoted here as MA-TCC, with window length $W\!=\!10$ is also given in Fig.~\ref{WMAaMANum} for comparison. It is noted that the MA-TCC fluctuates around its control limit whereas the OWMA-TCC goes beyond its control limit clearly. This phenomenon can be explained by \textit{Theorem \ref{ThmIFD0}}, which says that the introduced IFs are not guaranteed detectable by the MA-TCC($10$). Overall, their detailed IFD results are given in Fig.~\ref{DetectionResultsAllNum} with blue and red lines, respectively.

\begin{figure}
\begin{center}
\includegraphics[width=7.7cm]{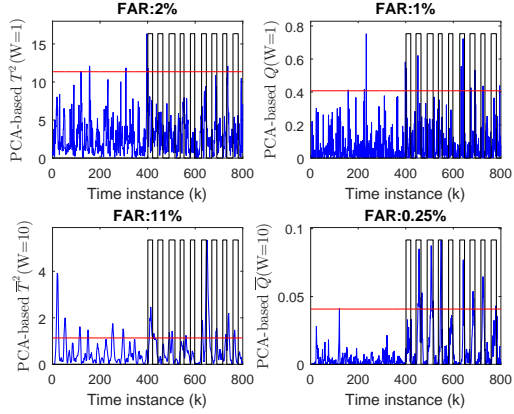}    
\caption{IFD using PCA-based and MA-PCA-based ($W\!=\!10$) control charts in the numerical simulation (Gaussian noise).}
\label{PCAaMAPCANum}
\end{center}
\end{figure}

Several static and dynamic MSPM methods are used here to show their limitations on dealing with IFs. Another 50000 consecutive observations are generated according to (\ref{NumericalModelY}) as training samples for these MSPM methods. The traditional PCA and its MA-based extension (i.e., the MA-PCA \cite{Ji2017Incipient}), are selected as the representatives of static MSPM methods.
For PCA and MA-PCA models, the cumulative percent variance (CPV) criterion says that three PCs should be chosen, which account for more than 95\% of the variance in original variables.
The MA-PCA-based $T^2$ and $Q$ statistics with window length $W\!=\!10$, denoted here as PCA-based $\bar{T}^2(10)$ and PCA-based $\bar{Q}(10)$, are utilized for comparison.
The PCA-based and MA-PCA-based control charts of the test data are given in Fig.~\ref{PCAaMAPCANum}. Moreover, their detailed IFD results are given in Fig.~\ref{DetectionResultsAllNum} with yellow and green lines, respectively.
It can be seen that traditional PCA is inefficient for IFs and the MA-PCA has an unacceptable high FAR (11\%). This high FAR is expected since several studies \cite{Guo2017Anaccelerated,Ku1995Disturbance,Kruger2007Improved} have already indicated that monitoring autocorrelated data using static MSPM methods tends to produce excessive false alarms.

\begin{figure}
\begin{center}
\includegraphics[width=7.7cm]{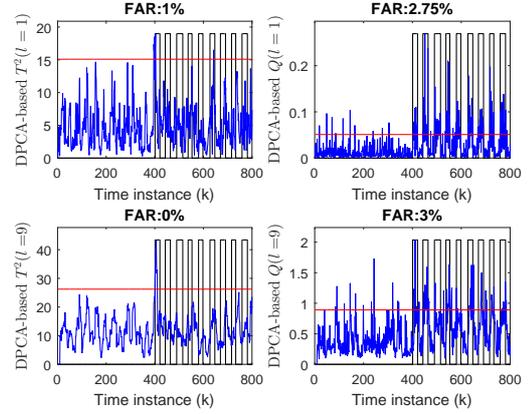}    
\caption{IFD using DPCA-based control charts with time lag $l=1$ and $l=9$ in the numerical simulation (Gaussian noise).}
\label{DPCAsNum}
\end{center}
\end{figure}

As for dynamic MSPM methods, we select DPCA \cite{Ku1995Disturbance} as their representative in this subsection, because the simulation model (\ref{NumericalModelY}) was first introduced therein. According to \cite{Ku1995Disturbance}, the time lag is determined as $l\!=\!1$, and five PCs are chosen for the DPCA model. The DPCA-based $T^2$ and $Q$ statistics of the test data are given in Fig.~\ref{DPCAsNum}. Moreover, their detailed IFD results are given in Fig.~\ref{DetectionResultsAllNum} with solid and dashed cyan lines, respectively. It is obvious that the IFD performance of both statistics is far from satisfactory.
For further comparison, the time lag is chosen as $l\!=\!9$, so that the same number of samples with OWMA-TCC(10), i.e., 10 samples, can be utilized to detect IFs at each time instance. According to the CPV criterion, twelve PCs should be chosen for the DPCA model at this time, which account for more than 99\% of the variance in original variables. The DPCA-based $T^2(l=9)$ and $Q(l=9)$ statistics of the test data are given in Fig.~\ref{DPCAsNum}, along with their detailed IFD results given in Fig.~\ref{DetectionResultsAllNum} with solid and dashed magenta lines, respectively. It can be seen that the IFD performance is still unsatisfactory.

\begin{figure}
\begin{center}
\includegraphics[width=7.7cm]{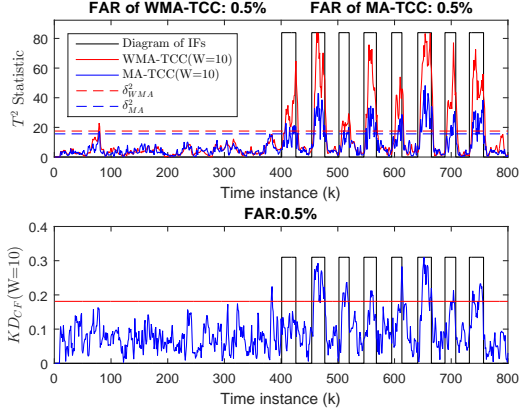}    
\caption{IFD using the OWMA-TCC, MA-TCC and MW-KD with window length $W\!=\!10$ in the numerical simulation (uniformly distributed noise).}
\label{WMAaMAaKDNumUN}
\end{center}
\end{figure}

\begin{figure}
\begin{center}
\includegraphics[width=7.7cm]{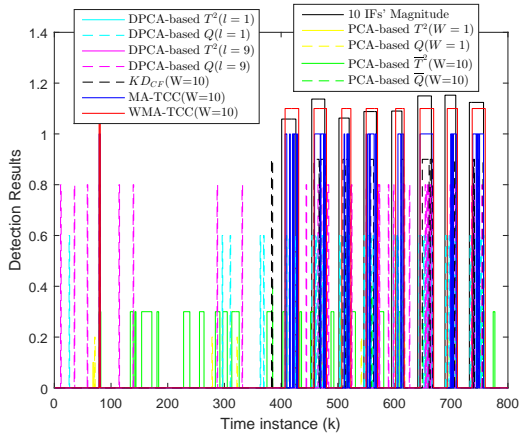}    
\caption{IFD results using different methods in the numerical simulation (uniformly distributed noise).}
\label{DetectionResultsAllNumUN}
\end{center}
\end{figure}

To demonstrate the effectiveness of the developed method under non-Gaussian conditions, the noises $\textbf{w}$ and $\textbf{v}$ are reset to uniform distributions ${\mathbb U}(-0.5,0.5)$ and $\sqrt{0.1}{\mathbb U}(-0.5,0.5)$, respectively. Except for this and resetting the lower bound of each fault magnitude to $\tilde{f}_q\!=\!0.105$, other parameters of the numerical example remain unchanged. A present MSPM method used for IFD under Gaussian or non-Gaussian condition, i.e., the MW-KD \cite{Kammammettu2019Change}, is employed. According to \cite{Kammammettu2019Change}, three PCs which account for more than 95\% of the original variance should be retained, and the threshold is set accordingly. The empirical method \cite{Shang2017Recursive} is used here to set the control limits of OWMA-TCC and MA-TCC. The statistics of OWMA-TCC($10$), MA-TCC($10$) and MW-KD($10$) are given in Fig.~\ref{WMAaMAaKDNumUN}. Their detailed IFD results are given in Fig.~\ref{DetectionResultsAllNumUN} with red, blue and dashed black lines, respectively. Moreover, IFD results of the above-mentioned static and dynamic MSPM methods are also given therein. By comparison, the importance of employing an optimal weight vector is observed.


To appreciate the performance of different methods, their IFD results under Gaussian and non-Gaussian conditions are shown together in Figs.~\ref{DetectionResultsAllNum} and \ref{DetectionResultsAllNumUN}, respectively. It is noted that only OWMA-TCC goes beyond its control limit clearly when an IF occurs, under both Gaussian and non-Gaussian conditions. By contrast, the others tend to fluctuate around their corresponding control limits. Thus, it can be seen that the developed method shows better IFD performance among several static and dynamic MSPM methods.

\subsection{The CSTR process}\label{CSTRSimuSub}
In this subsection, a continuous stirred tank reactor (CSTR) simulation is utilized to demonstrate the effectiveness and efficiency of the proposed methods through comparative studies.
The CSTR process can be described by the following differential equations
\begin{small}
\begin{align}\label{CSTRModel}
\frac{\mathrm{d}C_A}{\mathrm{d}t}&=\frac{q}{V}(C_{Af}-C_A)-k_0\exp\left(-\frac{E}{RT}\right)C_A+v_1,\\
\frac{\mathrm{d}T}{\mathrm{d}t}&=\frac{q}{V}(T_f-T)-\frac{\Delta H}{\rho C_p}k_0\exp\left(-\frac{E}{RT}\right)C_A+\frac{UA}{V\rho C_p}(T_c-T)+v_2,\nonumber
\end{align}
\end{small}
where $C_A, T, T_c, q, C_{Af}, T_f$ are the outlet concentration, reactor temperature, cooling water temperature, feed flow rate, feed concentration and feed temperature, respectively. $v_1$ and $v_2$ are independent Gaussian white noises. The measured variables are $[C_A, T, T_c, q]^\mathrm{T}$, where $[C_A, T]^\mathrm{T}$ are controlled variables with nominal values, and $[T_c, q]^\mathrm{T}$ are manipulated variables with feedback control.
More detailed descriptions of the CSTR process can be found in \cite{Li2010Reconstruction}, where the settings of the process, including system parameters and conditions as well as controller information, are also given therein. Different from most existing literature that always sets the sampling interval as 1min (in this situation, process data are nearly independent), we choose the sampling interval as 3s here because of the higher sampling frequency requirement for capturing IFs. Note that shortening the sampling interval results in autocorrelated process data.

\begin{figure}
\begin{center}
\includegraphics[width=7.7cm]{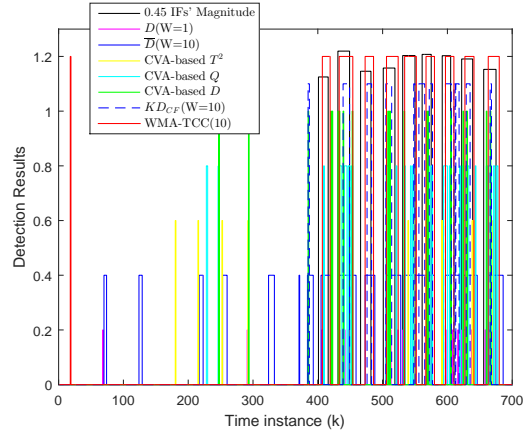}    
\caption{IFD results using different methods in the CSTR process (Gaussian noise).}
\label{DetectionResultsAllCSTR}
\end{center}
\end{figure}

The unmeasurable feed temperature $T_f$ is a main disturbance in the process, and has been used by many studies \cite{Shang2017Dominant,Shang2015Concurrent} to evaluate different FD methods. In this simulation, intermittent increases of feed temperature $T_f$ are introduced since sample 401, with a lower bound of each fault magnitude $\tilde{f}_q\!=\!2.5$K, a lower bound of each fault active, and inactive duration $\tilde{\tau}^o_q\!=\!\tilde{\tau}^r_{q}\!=\!10$ sampling intervals, i.e., 30s. The first 400 samples are used to calculate FARs of different methods. A total of 700 consecutive observations are collected as test samples. The actual fault magnitude, fault active and inactive duration are all generated randomly according to their lower bounds and are shown in Fig.~\ref{DetectionResultsAllCSTR} with a black line (the $Y$-axis shows the fault magnitude multiplied by 0.45, and the $X$-axis shows the fault active and inactive duration).

\begin{figure}
\begin{center}
\includegraphics[width=8.4cm]{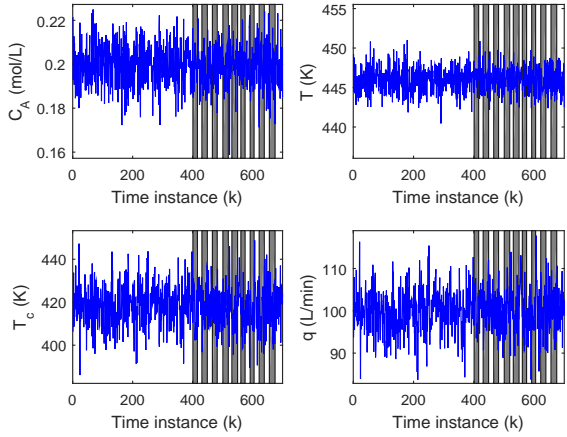}    
\caption{Measured CSTR process variables with intermittent disturbances in the feed temperature $T_f$ (Gaussian noise).}
\label{IFcstr}
\end{center}
\end{figure}

According to the process model \eqref{CSTRModel}, $T_f$ directly affects the reactor temperature $T$. However, since $T$ is controlled by manipulating the cooling water temperature $T_c$, when $T$ deviates from its nominal value, $T_c$ is immediately adjusted to compensate the change. In this way, the entire process is always under control, rendering the system parameters and conditions unchanged.
Therefore, when intermittent disturbances of $T_f$ occur, $C_A, T, q$ are still around their set-point values, whereas $T_c$ exhibits intermittent biases instead. This phenomenon is also shown in Fig.~\ref{IFcstr}, where collected process data with intermittent disturbances in $T_f$ are plotted and the gray shadows represent the active duration of IFs.
Moreover, note that the correlations (autocorrelation and cross-correlation) of process variables in this scenario remain unchanged. This can be seen from \eqref{CSTRModel} that time constants of $C_A, T$ are irrelevant with $T_f, T_c$.
As a result, the introduced intermittent disturbances in $T_f$ can be well modeled by (\ref{IFModel}) with fault direction $\xi_q\!=\![0,0,1,0]^T$.

Five thousand sets of 10 consecutive observations are collected under normal conditions as training samples, which are subsequently utilized to determine the optimal weight vector and the control limit with significance level $\alpha\!=\!0.01$. Then, we can conclude that the introduced intermittent disturbances in $T_f$ are guaranteed detectable by the OWMA-TCC with window length $W\!=\!10$, according to \textit{Theorems \ref{ThmIFD0}} and \ref{ThmIFD1} and \textit{Corollary \ref{ColIFDlow1}}.
Several well-known static and dynamic MSPM methods are also employed here for comparison. The Mahalanobis distance (MD) (also known as the global Hotelling's $T^2$ test $D$) \cite{Qin2003Statistical}, and its MA-based extension \cite{Ji2017Incipient} with window length $W\!=\!10$, i.e., $\bar{D}(10)$, are chosen as representatives of static MSPM methods. As for dynamic MSPM methods, CVA is chosen as their representative. Additionally, the MW-KD is also chosen. Another 50,000 consecutive observations are collected under normal conditions as training samples for these MSPM methods.
\begin{figure}
\begin{center}
\includegraphics[width=7.7cm]{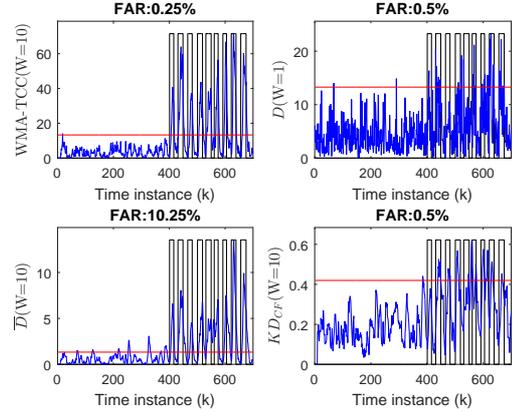}    
\caption{IFD using OWMA-TCC, MD, MA-MD and MW-KD in the CSTR process (Gaussian noise).}
\label{WMAaDaMADaKDCSTR}
\end{center}
\end{figure}

\begin{figure}
\begin{center}
\includegraphics[width=7.7cm]{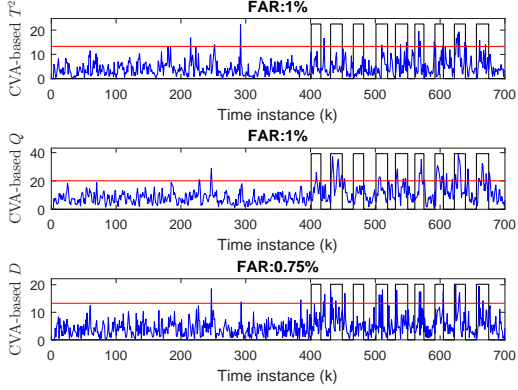}    
\caption{IFD using CVA-based control charts in the CSTR process (Gaussian noise).}
\label{CVACSTR}
\end{center}
\end{figure}

The OWMA-TCC($10$), $D(1)$, $\bar{D}(10)$ and $KD_{CF}(10)$ control charts of the test data are given in Fig.~\ref{WMAaDaMADaKDCSTR}. Moreover, their detailed IFD results are given in Fig.~\ref{DetectionResultsAllCSTR} with red, magenta, blue and dashed blue lines, respectively. For $KD_{CF}(10)$, two PCs which account for more than 95\% of the original variance are retained \cite{Kammammettu2019Change}. It can be seen that the $D(1)$ statistic is inefficient for IFs. While the traditional MA technique can indeed improve the statistics' sensitivity to IFs, it causes an unacceptable high FAR (10.25\%) when process data are autocorrelated, and consequently invalidates the online monitoring approach. By contrast, the proposed OWMA-TCC goes beyond its control limit clearly when IFs occur, and the FAR is consistent with its theoretical value, i.e., less than 1\%.
As for the CVA model, according to \cite{Pilario2018Canonical}, the number of time lags for past ($p$) and future ($f$) observations is determined using autocorrelation analysis on the training samples. For the simulation, it has been found that three time lags are the maximum, after which autocorrelations become insignificant for the summed squares of all measurements as well as for all the process variables, at 99\% confidence level. Thus, we set $p\!=\!f\!=\!3$. In addition, the number of states is chosen as four according to the dominant singular value (SV) method (to find the point where a ``knee" appears in the SV curve). The CVA-based $T^2, Q, D$ statistics \cite{Pilario2018Canonical} of the test data are given in Fig.~\ref{CVACSTR}, and their detailed IFD results are given in Fig.~\ref{DetectionResultsAllCSTR} with yellow, cyan and green lines, respectively. The IFD results indicate that CVA also has limitations on dealing with IFs. The time lags of CVA are chosen only based on system dynamics without taking the characteristics of IFs into account, resulting in a lack of sensitivity to IFs of the method.

\begin{figure}
\begin{center}
\includegraphics[width=7.7cm]{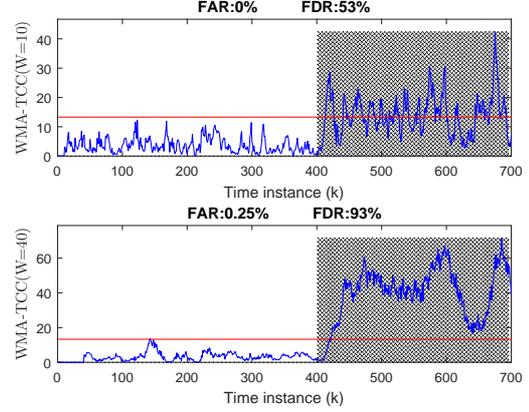}    
\caption{IFD using OWMA-TCC in the CSTR process ($\tau^o_q\!=\!\tau^r_q\!=\!1$, Gaussian noise).}
\label{WMAdur1CSTR}
\end{center}
\end{figure}

\begin{figure}
\begin{center}
\includegraphics[width=7.7cm]{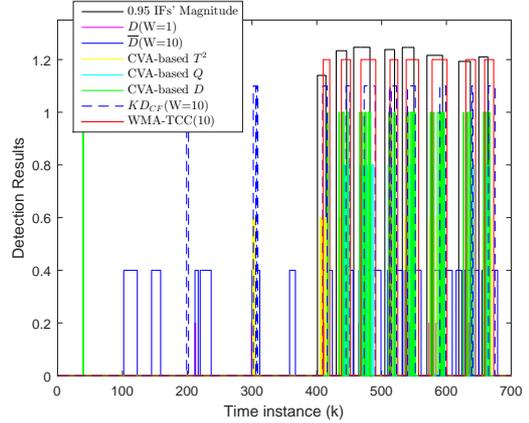}    
\caption{IFD results using different methods in the CSTR simulation (uniformly distributed noise).}
\label{DetectionResultsAllCSTRUN}
\end{center}
\end{figure}

In the case of $\tau^o_q\!=\!\tau^r_q\!=\!1$, the statistics of OWMA-TCC($10$) and OWMA-TCC($40$) are given in Fig.~\ref{WMAdur1CSTR}. It can be seen that due to the violation of $W\!\leq\!W^\#$, although the control chart still alarms, we can not determine each appearance (disappearance) of an IF before its subsequent disappearance (appearance). Moreover, due to the inclusion of both faulty and fault-free samples in the time window, the detectability condition is no more satisfied when $W=10$, resulting in the missed alarms. Nevertheless, the developed OWMA-TCC is still applicable in this case due to the efforts we have made in improving the existing MA-type schemes to smooth autocorrelated data. By enlarging the window length $W$ from $10$ to $40$, the missed alarms can be totally eliminated after some delay. In addition, to demonstrate the effectiveness of the developed method under non-Gaussian conditions, the noises $v_1$ and $v_2$ are reset to the uniform distribution $0.1{\mathbb U}(-0.5,0.5)$. Except for this and resetting the lower bound of each fault magnitude to $\tilde{f}_q\!=\!1.2$, other parameters of the simulated CSTR process remain unchanged. IFD results of the above-mentioned MSPM methods are given in Fig.~\ref{DetectionResultsAllCSTRUN}. By comparison, the better IFD performance of the OWMA-TCC is observed.

Finally, to appreciate the performance of different methods, their IFD results under Gaussian and non-Gaussian conditions are shown together in Figs.~\ref{DetectionResultsAllCSTR} and \ref{DetectionResultsAllCSTRUN}, respectively. It is noted that OWMA-TCC alarms continuously when an IF occurs. By contrast, the others tend to alarm sporadically, or start to alarm after the IF has disappeared. Overall, it can be seen that the developed method shows better IFD performance among the static and dynamic MSPM methods being compared.

\section{Conclusion and future perspective}\label{ConclusionSec}
In this paper, a weighted moving average (WMA) scheme has been combined with the Hotelling's $T^2$ statistic to form an optimally weighted MA (OWMA) $T^2$ control chart (OWMA-TCC) used in weakly stationary processes. Compared with static MSPM methods such as PCA, OWMA-TCC employs a time window and an optimal weight vector (OWV) to improve its detection capability for IFs that always manifest themselves as repeated small and short fluctuations.
Compared with traditional MA-type schemes such as MA-PCA, OWMA-TCC overcomes the problem of producing excessive false alarms when data exhibit autocorrelation, because it does not assume data to be independent. Moreover, OWMA-TCC can use the correlation (autocorrelation and cross-correlation) information to increase its sensitivity to IFs by finding an OWV.
Compared with dynamic MSPM methods such as DPCA and CVA, OWMA-TCC selects the window length considering the characteristics of IFs, i.e., the fault duration and magnitude, and then gains additional sensitivity to IFs by optimizing its weights.

The non-optimality of the equally and exponentially weighted scheme used for fault detection when data have autocorrelation has been discovered. The essence that existing MA-type schemes do not effectively utilize the correlation information of samples has been revealed. Then, an OWMA theory has been established, including methods to construct WMA statistics, analyze the fault detectability, and determine the OWV. Existence of the OWV has been proven with the help of the Brouwer fixed-point theory, and an iteration process to obtain the OWV has been provided. These ensure that the OWMA-TCC is implementable in real applications.
Moreover, we have found that the OWV possesses a symmetry structure, and the equally weighted scheme is optimal for any IF directions when data exhibit no autocorrelation. This verifies the optimality of existing MA-based MSPM methods when applied to independent data.
The developed method has been evaluated using a numerical example and the CSTR process. Simulation results have shown that for IFs with same direction, magnitude and duration, the compared methods, including several well-known static and dynamic MSPM methods, fail to detect them whereas OWMA-TCC succeeds in detecting them.

Further studies include the combination of OWMA with recursive methods, other statistics, kernel methods, dynamic data modeling methods and other selection criteria, to address the problems of monitoring processes with slightly varying operation points, varying levels of noise, nonlinear properties and nonstationary properties, as well as detecting faults with unknown characteristics.





\bibliographystyle{unsrt}
\bibliography{IFDWMATCC}







\end{document}